# 0D-2D Heterostructure for making very Large Quantum Registers using 'itinerant' Bose-Einstein Condensate of Excitons

Amit Bhunia[a], Mohit Kumar Singh[a], Maryam Al Huwayz[b,c], Mohamed Henini[b] and Shouvik Datta[a] *

[a] *Department of Physics, Indian Institute of Science Education and Research, Pune 411008, Maharashtra, India*
[b] *School of Physics and Astronomy, University of Nottingham, Nottingham NG7 2RD, UK,*
[c] *Physics Department, Faculty of science, Princess Nourah Bint Abdulrahman University, Riyadh, Saudi Arabia.*

***Email:*** shouvik@iiserpune.ac.in, Ppzmh@exmail.nottingham.ac.uk

# ABSTRACT

Presence of coherent 'resonant' tunneling in quantum dot (zero-dimensional) - quantum well (two-dimensional) heterostructure is necessary to explain the collective oscillations of average electrical polarization of excitonic dipoles over a macroscopically large area. This was measured using photo excited capacitance as a function of applied voltage bias. Resonant tunneling in this heterostructure definitely requires momentum space narrowing of charge carriers inside the quantum well and that of associated indirect excitons, which indicates bias dependent 'itinerant' Bose-Einstein condensation of excitons. Observation of periodic variations in negative quantum capacitance points to in-plane coulomb correlations mediated by long range spatial ordering of indirect, dipolar excitons. Enhanced contrast of quantum interference beats of excitonic polarization waves even under white light and observed Rabi oscillations over a macroscopically large area also support the presence of density driven excitonic condensation having long range order. Periodic presence (absence) of splitting of excitonic peaks in photocapacitance spectra even demonstrate collective coupling (decoupling) between energy levels of the quantum well and quantum dots with applied biases, which can potentially be used for quantum gate operations. All these observations point to experimental control of macroscopically large, quantum state of a two-component Bose-Einstein condensate of excitons in this quantum dot - quantum well heterostructure. Therefore, in principle, millions of two-level excitonic qubits can be intertwined to fabricate large quantum registers using such hybrid heterostructure by controlling the local electric fields and also by varying photoexcitation intensities of overlapping light spots.

## 1. Introduction

Experimental detection of macroscopically large, phase coherent quantum state within semiconductor heterostructures can always open up new paradigms for emergent quantum phenomena in material science. In this study, we will show how excitons [1] which are quasiparticles of bound electron-hole pairs can produce these experimentally tunable, two-level quantum states using a quantum coupled, zero dimensional (0D) - two dimensional (2D) heterostructure. To begin with, we detect experimental signatures of phase coherent oscillations of average electrical polarization of excitonic dipoles using large area capacitance measured under photo excitations or photocapacitance to probe this quantum phenomenon. We used Molecular beam epitaxy to grow quantum dots (QDs) based single crystal sample of III-V semiconductors as a double-barrier, resonant tunneling diode of p-GaAs/AlAs/InAs/AlAs/n-GaAs to investigate all these cooperative, many body effects of excitons towards their proposed application as qubits to build very large quantum registers. The 'Materials and methods section' at the end will provide all relevant details of growth of this 0D-2D heterostructure, subsequent device fabrications and experimental methods used to measure photocapacitance signals from excitons.

In general, coherent and incoherent tunneling of electrons and holes were thoroughly explored [2-4] earlier. Multiple peaks [5] in photocapacitance as charging of QDs [6] and correlations of excitonic complexes [7] were studied before. Quantum oscillations of photocurrents in this coupled 0D-2D heterostructure were also investigated in the past [8-10]. These were some early observations of 'quantum' oscillations in QD based resonant tunneling diodes. However, none of those studies could explain – (a) why so many of these electrons/holes take part 'collectively' in coherent resonant tunneling processes across millions of QDs spread over hundreds of microns to produce phase coherent oscillations in both dc and ac measurements and

(b) what exactly is 'quantum' in these so called quantum oscillations of photocurrent beyond a nominal temperature independence of oscillation peaks as discussed [8] in the past. So in section 2, we will argue why 'coherent' nature of resonant tunneling are necessary to explain uncharacteristically pronounced oscillations due to tunneling of heavier holes in forward bias, which was not at all addressed earlier. Next we will present a series of experimental evidences to demonstrate the presence of inherently 'quantum' phenomena behind these phase coherent, photocapacitance oscillations.

In section 3, we will discuss the direct role of excitons in generating these photocapacitance oscillations. Further observations of negative quantum capacitance, in section 4, will point to positional correlations among charge carriers within the 2D quantum well (QW) at first and then consequently among the associated 0D-2D spatially indirect excitons. These negative quantum capacitance induce coulomb correlations were mostly reported [11-13] in case of 2D-2D bi-layer electron systems in the past. However, here we will extend those interpretations (Ref. 11, Fig. 5 ibid.) and apply the same directly to this 0D-2D excitonic system.

In the following sections 5 and 6, we will argue why experimental observations of 'quantum' oscillations associated with collective undulations of electrical polarization of these dipolar excitons indicate the presence of an 'itinerant' Bose-Einstein condensation (BEC) of excitons in this 0D-2D heterostructure from several perspectives. To support this particular understanding, we will then present a sequence of experimental findings and related analyses. These include – (a) an unusual increase in magnitude of photocapacitance oscillation signifying spontaneous enhancement of collective electrical polarization of dipolar excitons below ~100 Kelvin in section 5, (b) why observation of coherent resonant tunneling in this 0D-2D heterostructure requires momentum space narrowing which indicates the presence of excitonic

BEC in section 6, (c) the presence of long range correlations in the form of phase coherent interference of electric polarization of excitons photo generated even with incoherent white light as evidence of density driven excitonic order in section 7, (d) presence of Rabi oscillations of macroscopic, two-level quantum state of excitonic BEC as a function of photoexcitation intensity in section 8. Additionally, we will present observations of periodic presence (absence) of splitting of excitonic peaks in photocapacitance spectra to argue quantum coupling (decoupling) between energy levels of the quantum well and quantum dots with applied biases in section 9. We will show that such 'quantum coupling-decoupling' can potentially be used for Hadamard like quantum gate operations. Section 10 will further discuss how one can experimentally tune the Bloch sphere of such two-level quantum state of excitonic BEC with both applied voltage bias and photoexcitation intensity. Then we will also argue why photocapacitance is uniquely capable to detect these macroscopically large, multi-partite quantum state of collective electrical polarization of dipolar excitons in section 11. Section 12 will discuss why the observed long range quantum coherence can persist even in presence of tunneling.

Based on these observations, finally, in section 13, we will propose that experimental control of macroscopic quantum state of this two-component excitonic BEC in such 0D-2D heterostructure can be used to fabricate quantum register of qubits consisting of millions of bosonic clones in the form of excitons. In section 14, we will provide summary of our studies in the context of quantum device applications. There we will also highlight the importance of further research in fabricating such 0D-2D heterostructures using materials having higher excitonic binding energies to probe and utilize these phase coherent phenomena of excitons.

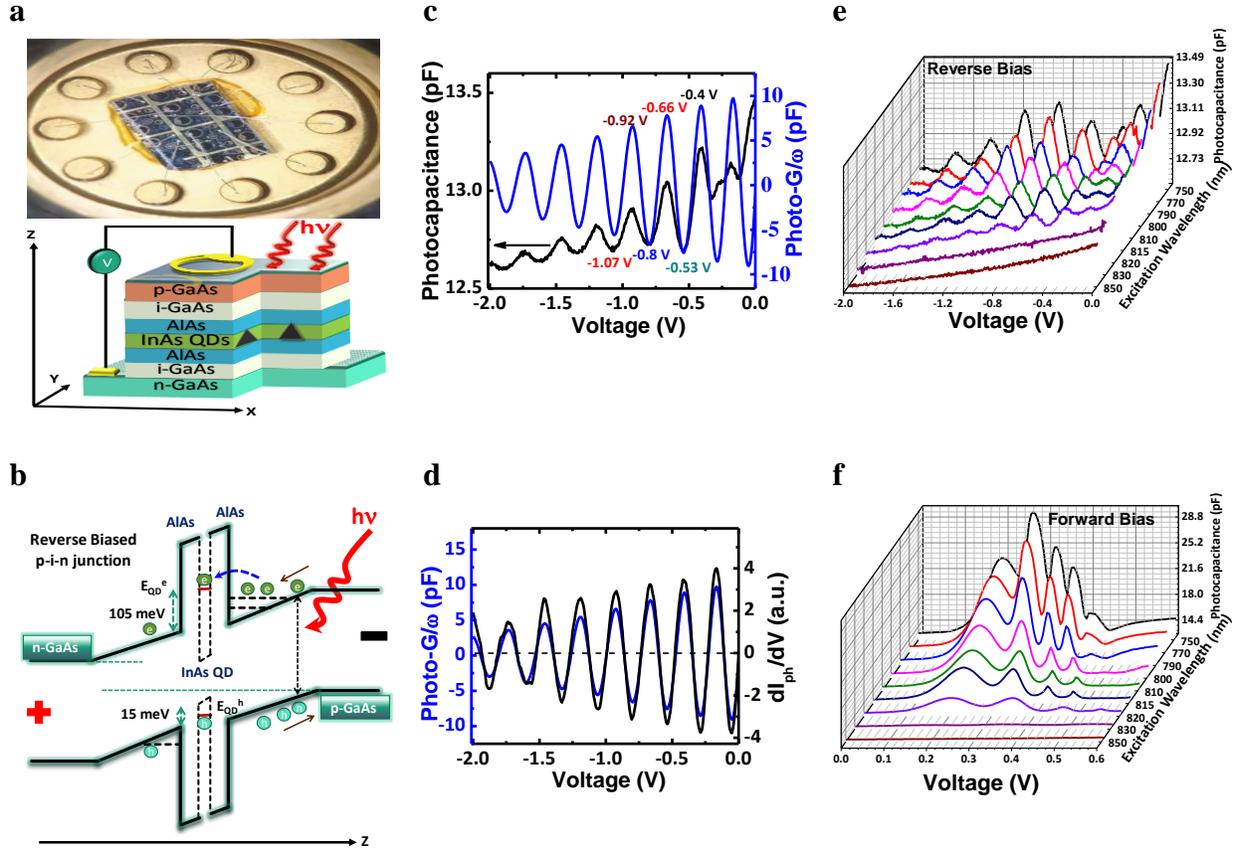

**Fig. 1.** Sample structure, band diagram & photocapacitance oscillations. (a) Schematic diagram of GaAs/AlAs/InAs/AlAs/GaAs double-barrier heterostructure below the actual image of the sample. Black triangles represent the InAs QDs sandwiched between two AlAs layers. Regions between the QDs are filled with AlAs capping layer and InAs wetting layer, which we do not explicitly show in this diagram. (b) Corresponding energy band-diagram with a sizable 2D electron accumulation layer formed mostly in the top, illuminated side of the AlAs barrier under reverse bias. Oval shape indicates the formation of 0D-2D indirect exciton under reverse bias. Brown arrows show the drift direction of electrons and holes, '+' and '-' signs indicate the direction of applied reverse bias and $\hat{z}$ is the epitaxial growth direction of this heterostructure. Blue arrow represents coherent tunneling of electrons. $E_{QD}^e$ & $E_{QD}^h$ are ground state energy levels of electrons and holes within InAs QD respectively. (c) 'Quantum' oscillations of photocapacitance and photo-G/ω for photoexcitation with red light (~ 630 nm using a halogen lamp and a monochromator), ω is the angular frequency of 10 kHz. (d) Photo-G/ω and $dI_{Ph}/dV$ oscillations with respect to bias. (e) and (f) Photocapacitance oscillations under wavelength selective photo excitations using the monochromator for both reverse and forward biases respectively.

## 2. Observations of Phase Coherent Collective Oscillations of Excitonic Dipoles

Schematic diagram of the sample is given in Fig. 1a. Energy band diagram under reverse (forward) bias is strongly affected by the accumulation of photo generated electrons and holes and is illustrated in Fig. 1b. A two-dimensional electron gas (2DEG) can form inside a GaAs triangular QW (TQW) whenever the quasi Fermi level for accumulated electrons on the p-type side of AlAs barrier crosses the GaAs conduction band edge under reverse bias. It is expected that in the steady state, some of these accumulated holes on the other side of AlAs barrier can reside in InAs QDs under finite biases. Thus indirect excitons are formed, where electrons (within GaAs 2DEG) and holes (within InAs QD) in two different layers are spatially separated by AlAs potential barrier along the growth direction ($\hat{z}$) of the sample as shown in Fig. 1b. One expects larger photo generations in top, illuminated p-type side of AlAs as compared to that in the lower n-type side. As a result, a sizably larger 2DEG can accumulate in the top p-type side under reverse bias and we ignore relatively smaller accumulation of holes on the n-type side of AlAs barrier. Here we also ignore the QD wetting layers. We measured photocapacitance [14-16] using ~200 μm wide ring shaped electrical top-contact covering ~$10^{11}$/cm$^2$ of InAs QDs. Most importantly, it was known [14,15], that photocapacitance is proportional to orientational average of electric polarization of excitonic dipoles in such structures. Dipole moments of such indirect excitons were estimated and compared with those of indirect excitons in the past [14,15]. Further information about the heterostructure growth and experimental methods can be found in the 'Materials and methods section' at the end. It is worth noting that any usual band diagram with a depleted junction in such a p-i-n diode heterostructure having a potential barrier in the middle will not work here without considering the applied bias and accumulated charge carriers near AlAs (Ref. 15, Fig. 3, ibid.).

In Fig. 1c, we show photocapacitance and photo conductance (photo-G/ω) oscillations versus bias ($V$) measured at 10 K. In Fig. 1d, we find that photo-G/ω and $dI_{Ph}/dV$ oscillations coincide, where $I_{Ph}$ is the DC-photo current. However, photo generation of currents ($I_{Ph}$) anywhere in the sample structure also affect photo-G/ω. Incidentally, at this light intensity, maxima (minima) of photocapacitance almost coincide with maxima (minima) of tunneling conductance as photo-G/ω or $dI_{Ph}/dV$ at the inflection points of $I_{Ph}$. So the maxima of $I_{Ph}$ are not necessarily maximizing bias driven accumulation of excitonic dipoles around the 0D-2D heterojunction which is expected in any sequential, incoherent tunneling process in steady state [8-10].

We now try to understand whether resonant tunneling of electrons across the AlAs barrier between 2D GaAs TQW and millions of 0D InAs QDs are either coherent or incoherent. It was argued earlier [8] that these oscillations can happen via three different ways: (i) sequential electron emissions [8] from successive bound states of 2DEG in the GaAs TQW towards the p-type side, (ii) repeated, non-coherent, sequential resonant tunneling between the 1$^{st}$ TQW level and 1$^{st}$ InAs QD level and (iii) periodic emptying of successive quantum states of GaAs TQW due to non-coherent resonant electron tunneling from GaAs 2DEG to InAs QDs with increasing reverse bias. However, it is unlikely that electron Fermi levels will be situated near the top edge of the GaAs TQW barrier for this 2DEG, which can contribute to successive electron escape towards the p type side with increasing reverse bias. Moreover, we expect more electron and hole accumulations around the AlAs barriers and subsequently a deeper TQWs with increasing biases as well as with increasing photo excitations. A deeper TQW and successive bound to unbound transition of quantized TQW levels with electrons escaping from higher levels towards top p-type side would have resulted in decreasing bias gaps of successive oscillation peaks with both increasing biases and also with increasing photo excitations. All these are neither observed in Fig. 1 or in the past [8-10]. Options (ii) and (iii) would have also shown similar progressive reduction of observed periodicity of photocapacitance/photocurrent oscillations with higher reverse biases. Therefore, we rule out all three possibilities of sequential and incoherent tunneling and/or generically incoherent diffusion [8-10] processes from our considerations.

As more and more photo generated charge carriers, under increasing applied electric field or bias, are driven towards AlAs, it is expected that tunneling contribution towards these oscillations should also increase with enhanced accumulations. However, magnitude of photocapacitance oscillations actually decreases (Fig. 1c) at both higher reverse and forward biases. Therefore, any sequential, incoherent [8-10] tunneling processes through millions of InAs QDs as driving mechanism for these oscillations are definitely not adequate explanations. Instead, we will demonstrate how coherent tunneling can allow the long range phase coherence to persist in these measurements. The observed phase coherence over such a large area can only decrease in presence of increasing number of bias driven, photo generated charge carriers. Subsequently these oscillations also decay at higher biases (Fig. 1c). Presence of periodically negative regions of photo-G/ω further revealed the absence of any simple-minded charging/discharging of QDs as the mechanism for such oscillations as well.

In order to understand why 'coherent' tunneling is certainly required to explain all these oscillations, we need to compare the nature of oscillations under both forward and reverse bias. Normally, photo generated holes will drift and accumulate near the illuminated side of the AlAs barrier under forward bias. As a result, a similar band alignment with inversed tilt [14,15] is produced under forward bias due to significant accumulation of holes in the top, illuminated p-GaAs side of AlAs barrier. A two-dimensional hole gas (2DHG) forms when the quasi Fermi level for holes accumulated on the p-type side of AlAs barrier crosses the GaAs valence band edge under forward bias. Tunneling rate of heavier holes are expected to be much less compared to lighter electrons. However, most interestingly, photocapacitance oscillations are less pronounced and more widely spaced in reverse bias (Fig. 1e) as compared to forward bias (Fig. 1f). Moreover, these oscillations survive even when excited below 1.61 eV (~band gap of InAs QDs), but fully subside for photo excitations below GaAs bandgap at 10 K. It is also interesting to see that maxima and minima of photocapacitance oscillations are shifting in both Figs. 1e, 1f with photo excitation wavelengths as well. Although, we agree that this closely spaced oscillations under forward bias originate from closely spaced hole energy levels in TQW due to a higher value of effective mass of holes as compared to electrons. However, these pronounced oscillation magnitudes in forward biases cannot [8-10] be explained using any simplified model of incoherent resonant tunneling of holes. Moreover, Figs. 1e, 1f even show that the position of maxima and minima of photocapacitance oscillations with applied bias is not firmly associated with energy level matching at one particular bias for any sequential incoherent resonant tunneling [8-10] of charge carriers but depends on photoexcitation intensity and wavelength as well. Nevertheless, it was known that peak-to-valley ratio of Fabry-Perot oscillations of charge carriers during coherent resonant tunneling [17] in a double barrier structure can be larger if barrier transmission probabilities are smaller. Therefore, coherent tunneling is necessarily required to explain the observed enhancements of phase coherent oscillations under forward bias as compared to reverse bias, which non-coherent electron tunneling models [8-10] cannot explain at all.

## 3. Role of Excitons in these Oscillations

In Figs. 2a and 2b, we plot photocapacitance spectra measured at 0.326 V and 0.6 V respectively. The 0.326 V bias is located at one of the peak position within the oscillation regime from 0.0 V to 0.5 V, whereas the 0.6 V lies beyond the oscillation regime observed in of Fig. 1f. It is important to note that we see nice resonant peak like excitonic features whenever photocapacitance spectra are measured inside the oscillation regime but there is no trace of excitonic presence once the spectra are measured beyond the 0.5 V. Another way to interpret this observation is to argue that presence of excitons in this 0D-2D heterostructure is linked with these oscillations in photocapacitance versus voltage measurements. We hardly see any pronounced excitonic peaks in Fig. 2c, whenever no external bias is applied in closed circuit condition under similar spectroscopic resolutions for photocapacitance spectra measured at 10 kHz. So it is clear that excitons are certainly not forming in significant numbers within this heterostructure in steady state in the absence of any applied voltage bias. The resonant peak in Fig. 2a is only due to bias driven accumulations of photo generated charge carriers around AlAs barrier and resulting formation of 0D-2D indirect excitons. It indicates the crucial role of these photo excited and bias driven formation of indirect excitons in connection with observing these oscillations. We will further probe these connections in subsequent sections. Specifically, intricate variations in excitonic photocapacitance spectra at bias values of maximum and minimum of photocapacitance oscillations will also lead us to understand the role of double-well like quantum coupling and uncoupling of 0D InAs QDs and 2D GaAs reservoirs, which will be discussed in section 9.

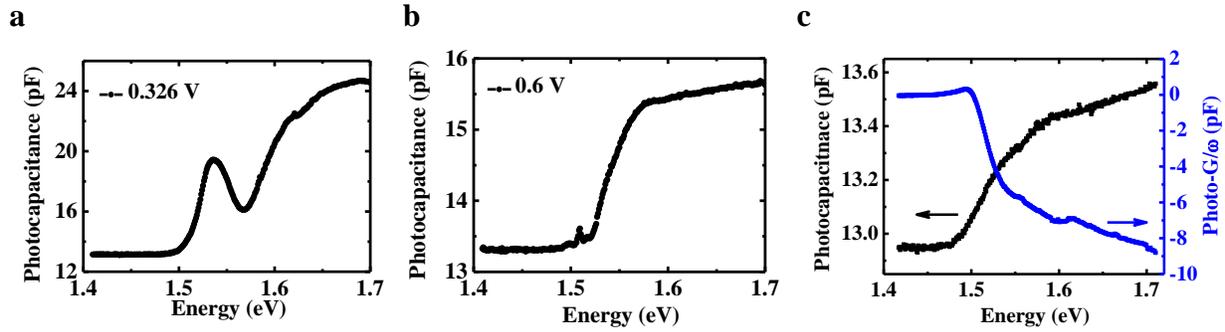

**Fig. 2.** Role of excitons in photocapacitance oscillations. (a) and (b) show photocapacitance spectra measured at two different forward bias values of 0.326 V and 0.6 V, respectively. We observe a prominent excitonic peak at 0.326 V, whereas we do not see clear excitonic peak at 0.6 V. However, we have seen earlier (Fig. 1f) that photocapacitance oscillation vanishes after 0.5 V. So, these observations apparently suggest that the presence of excitons are somehow related to the observed oscillations in photocapacitance as a function of voltage. (c) Excitonic resonances are barely visible under no applied bias.

**4. Negative Quantum Capacitance and Oscillations of Coulomb Correlated Excitons**

Oscillations of 'negative' quantum capacitance ($C_q$) [11-13] under reverse bias are also observed in Fig. 3. Here we assume that estimated quantum capacitance, due to coulomb screening effect of accumulated electrons within 2DEG, is in series with the dark capacitance to produce the measured photocapacitance. Please note that this dark capacitance does not show any such oscillations as a function of applied voltage bias as formation of 0D-2D excitons and related oscillations are mostly photo generated in our case. Here holes confined within InAs QDs attract electrons inside the 2DEG to create indirect excitons which are localized around these QDs. These dipolar excitons repel each other in the x-y plane as well. In fact, formation of these indirect excitons yield negative $\left(\frac{d\mu}{dn}\right)$ and subsequently generate the 'negative' quantum capacitance as

$$1/C_q \sim \left(\frac{d\mu}{dn}\right) \quad (1)$$

where $n$ is the number of indirect excitons and μ is excitonic chemical potential such that $na_B^2 < 1$ as well as $nd^2 < 1$, where $n \sim 10^{11}/cm^2$ as estimated below, $a_B$ (Bohr radius) for electrons in GaAs as 11.6 nm and d is the length of these indirect excitonic dipoles ~5 nm to 65 nm as the width of the undoped region on p-type side. As such surface charge density of indirect excitons ($n \sim \sigma_{ph}$) can be calculated [14,15] from $\sigma_{ph} e = CV = \langle \vec{P} \rangle \cdot \hat{z}$, where $e$ is the electronic charge and $C$ is photocapacitance per unit area at each bias $V$ and $\langle \vec{P} \rangle$ is the orientational average of electric polarization vector of these indirect excitonic dipoles. We estimate this $\sigma_{ph} \sim 1.099 \times 10^{11}/cm^2$ at 10.5 K. Following the theoretical predictions [11], one then finds that this 'negative' quantum capacitance can actually produce 'checkerboard' [11] like positional coulomb correlations between electrons of the 2DEG as a 2D Wigner lattice formed over these InAs QDs. See Ref. 11, schematic in Fig. 5 ibid. to understand this coulomb correlation effects in generating such formations. Here we invoke a similar qualitative picture for these coulomb correlated 0D-2D excitons in the x-y plane. As mentioned above, the coulomb correlated 2D electrons pair up with holes confined in the InAs QDs to form excitons. Generation of these indirect excitons further lower the overall potential energy of this excitonic system. So indirect excitons are formed at the minima of these Coulomb potential valleys ($\Phi^{EX}_{xy}$) localized near the InAs QDs. When more electrons accumulate in 2DEG with increasing reverse bias, then the in-plane coulomb correlation of these electrons get perturbed. Subsequently, a similar fate also awaits the associated 0D-2D indirect excitons. Therefore, increasing bias and more charge accumulations periodically drives these excitons away from the potential energy minima of in-plane interaction energy ($\Phi^{EX}_{xy}$) as a consequence of repulsive Coulomb interactions among the 2DEG electrons. Thereafter excitonic system again comes back to the minima of $\Phi^{EX}_{xy}$ when these excess 2DEG electrons tunnel through the 0D-2D heterojunction. In consequence of this many-body cooperative effect in the x-y plane, in addition to tunneling, these 2DEG electrons 'collectively' organize and reorganize in the Coulomb potential valleys ($\varphi^e_{xy}(V)$) around each InAs QDs. Consequently, the orientational average of electric polarization vector ($\langle \vec{P} \rangle$) of indirect excitonic dipoles associated with these 2DEG electrons also oscillate to produce the experimentally observed 'quantum' oscillation of photocapacitance as a function of applied bias over a macroscopically large area. As a result, the above mentioned self-organization of indirect excitons periodically 'freeze' around the InAs QDs in the x-y plane of the heterojunction and subsequently melt away with increasing bias voltages.

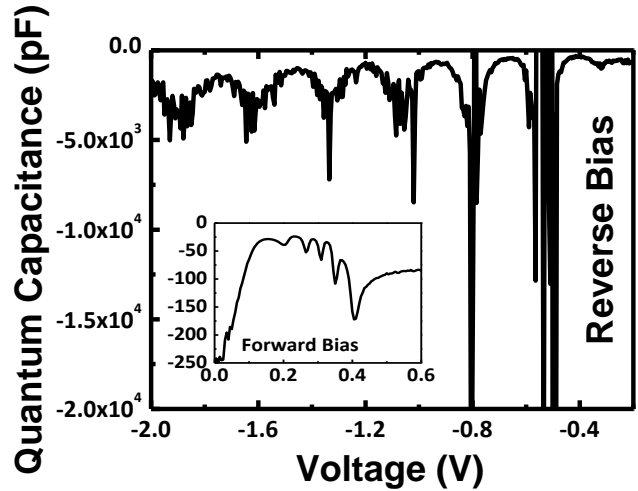

**Fig. 3.** Quantum capacitance of exctions. Oscillations of 'negative' quantum capacitance under reverse bias at 10 K. This indicate strong positional correlation of 2DEG electrons of these 0D-2D indirect excitons in the x-y plane. Similar quantum oscillations of smaller magnitudes are also observed as a function of forward bias (inset).

Being heavier and less mobile, holes have more potential energy of Coulomb correlation ($\varphi^h_{xy}(V)$) than their kinetic energy. Thus holes form a much 'stiffer' lattice of indirect excitons in the x-y plane. It is similar to having a much larger equivalent spring constant for oscillations as compared with lighter electrons under reverse bias. That also explains a smaller negative quantum capacitance (inset of Fig. 3) for holes and why measured photocapacitance oscillations are more pronounced and more closely spaced under forward bias as compared to reverse bias (Figs. 1e and 1f). This kind of understanding was missing in previous reports [8-10] of quantum oscillations. In earlier works [11-13], Landau quantization under strong magnetic field lead to localization of electrons. Similarly, in our case, formation of 0D-2D indirect excitons localize these 2DEG electrons around the InAs QDs. However, weak oscillations of photocurrents were also observed [8] in samples without InAs QDs only under forward biases whenever holes actually accumulate in the top p-GaAs layer near the AlAs barrier. So we guess that holes having more potential energy of Coulomb correlation can still localize these indirect excitons even in the absence of QDs as well.

Based on these understandings, we speculate that such tendencies to form 'Wigner crystallization' like periodically frozen, spatially ordered state of 2D electrons around the InAs QDs can actually serve as a precursor for excitonic BEC and vice versa by spontaneously enhancing the $(\langle \vec{P} \rangle)$ associated indirect excitons at periodic bias intervals. So, in principle, there could also be 'itinerant' formation of Wigner super-solid [18] if these excitons also undergo 'itinerant' BEC as a function of applied bias. However, in practice, spatial arrangements of these self-assembled InAs QDs are not exactly periodic. Therefore, any precise experimental evidence of any translational symmetry breaking in the x-y plane using photocapacitance measurement can be somewhat intricate and currently beyond the scope of this letter. Nevertheless, this realization certainly opens up further research avenues which can probe samples having increasingly more periodic or regular arrangements of QDs in the x-y plane of such 0D-2D heterostructures in future.

Until now we are able to establish the critical roles of excitons in these photocapacitance oscillations, coherent nature of resonant tunneling and exciton-exciton correlations based on negative quantum capacitance. Next, we will discuss sudden, unusual enhancement of oscillation magnitudes at low temperatures and why precise momentum matching will be required for any such resonant tunneling in this 0D-2D heterostructure and how that point towards bias dependent, itinerant excitonic BEC. It will turn out that unlike photocurrent, photocapacitance also oscillates with increasing light intensities as well (section 8).

## 5. Experimental evidence of spontaneous and collective enhancement of electric polarization of excitonic dipoles at low temperatures.

In this section, we will demonstrate spontaneous enhancements of collective electrical polarization $\langle \vec{P} \rangle$ of excitonic dipoles measured using photocapacitance over a large area. Some of the early reports of excitonic BECs were mostly using planar structures [19-24]. However, purely optical signatures of excitonic BEC are still being debated [25,26] as dark excitons [27,28] are expected to form the usual BEC ground state in most semiconductors. However, photocapacitance [14,15] was known to probe orientational average of electric polarization of excitonic dipoles irrespective of their dark or bright nature. Dipole moments of indirect excitons in similar structures were determined [14,15] from capacitance data and successfully compared with theoretical estimates for indirect excitons.

As such, thermally activated capacitive responses usually freeze with decreasing temperature. However, here in Figs. 4a and 4b, we observed atypical enhancements of both photocapacitance and photo-G/ω oscillations as more and more excitons begins to form at temperatures below ~100 K. Although peak positions shift with changing temperatures but the period of these oscillations is somewhat independent of temperature. Earlier, only this was suggested [8] as signs of 'quantum' nature of these oscillations. However, in this study, we will provide several other experimental observations to reveal the 'quantum' origins of this oscillations in subsequent sections.

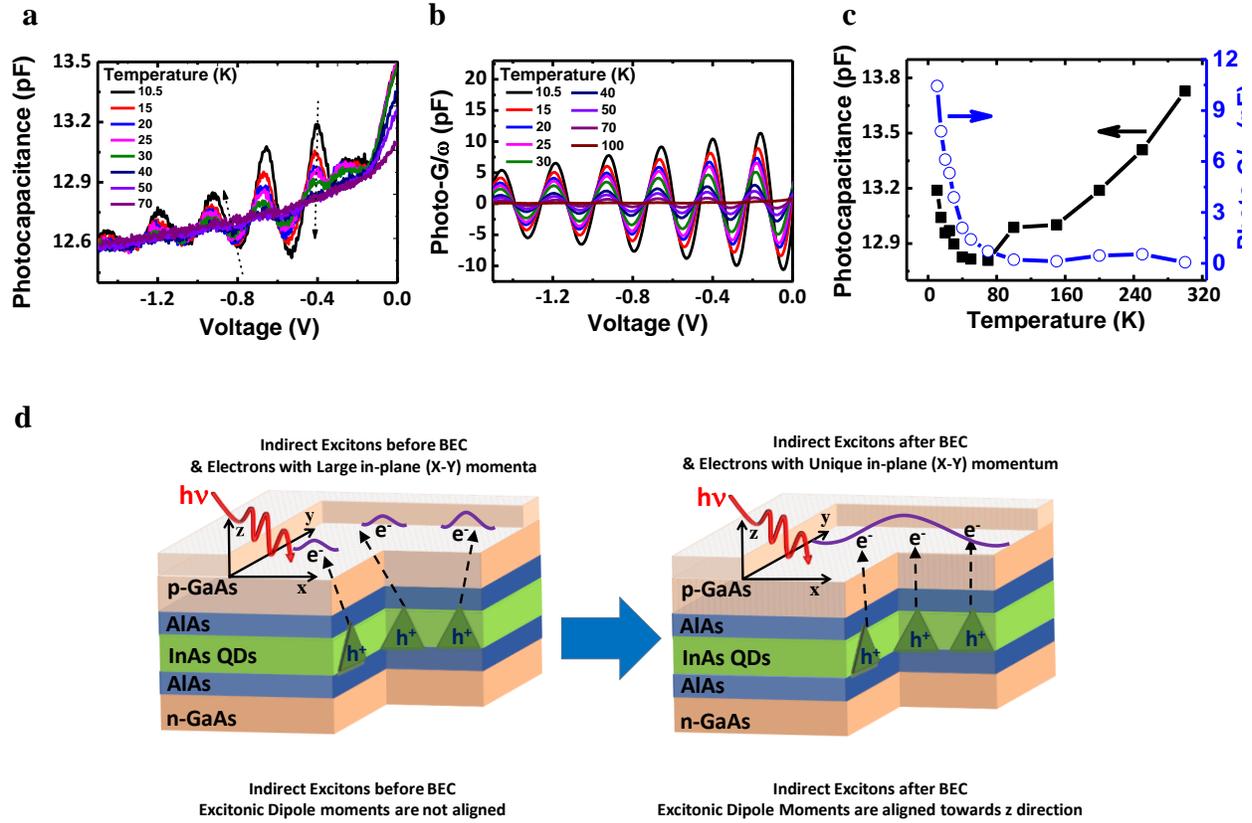

**Fig. 4.** Spontaneous enhancement of photocapacitance oscillation at low temperatures. (a) Photocapacitance and (b) Photo-G/ω oscillations at 10 kHz for different temperatures under selective photoexcitation centered around 630 nm using a halogen lamp. (c) Variation of peak oscillation magnitudes at -0.4 V are shown as a function of temperature. (d) Schematics of spontaneous enhancement of electric polarization $\langle\vec{P}\rangle$ of indirect excitons before and after excitonic dipoles collectively align along z direction. The purple contours on the left represent the wave functions of individual indirect excitons before the onset of BEC. The large wave function on the right 'roughly' represents the macroscopic quantum state of excitonic BEC once it is formed.

Fig. 4c depicts the unusual increase of maximum $\langle\vec{P}\rangle$ or $\sigma_{Ph}$ from ~$1.067\times10^{11}$/cm$^2$ around 70 K to ~$1.099\times10^{11}$/cm$^2$ at 10.5 K. Even the contrast $\left[\sim\frac{(C_{Max}-C_{Min})}{(C_{Max}+C_{Min})}\right]$ of these phase coherent oscillations as a measure of first order spatial correlation of $\langle\vec{P}\rangle$ of excitonic dipoles, measured over an area of diameter ~200 μm, suddenly increases with decreasing temperature. This likely indicate a continuous 2$^{nd}$ order phase transition in the excitonic density being probed with photocapacitance as $C\sim\langle\vec{P}\rangle$. See section 7 for further details of this interpretation. Discrete energy levels can survive in both of these 0D InAs QDs and 2DEG in p-GaAs accumulation layer [15] even at room temperatures. However, these oscillations wash out above ~ 100 K due to thermal effects [15] of energy level broadening, phonon scatterings and enhanced thermal activation and/or dissociation of excitons etc.

## 6. Why the presence of Coherent Resonant Tunneling in a 0D-2D Heterostructure indicates Excitonic BEC?

We now explore whether these periodic enhancements of $C \sim \langle \vec{P} \rangle$ are associated with excitonic BEC. Any 'coherent' resonant tunneling [17] of electrons of these indirect excitons 'necessarily' requires [29] strict energy and momentum matching conditions. Following the Fig. 1 of Ref. 29, we can see that for resonant tunneling in a 0D-2D system under reverse bias, one rquires -

$$E^e_{QD}(V,I) = E^e_{2DEG}(V,I) + \left[\frac{\hbar^2}{2m^*}(\vec{k}_x^2 + \vec{k}_y^2) + \Phi^{EX}{}_{xy}(V,I)\right] + (eV) \qquad (2)$$

where $E^e_{QD}$, $E^e_{2DEG}$ are respective ground-state energies, $\vec{k}_x$, $\vec{k}_y$ are in-plane momentum of 2DEG electrons, $\Phi^{EX}{}_{xy}(V,I)$ accounts for net Coulomb interaction (correlation) energy of these 0D-2D correlated [11] indirect excitons in the x-y plane and $I$ is the light intensity used for photoexcitation. Therefore, resonant tunneling at bias $V(z) = V_0, I = I_0$ is possible, only and only if the term $\left[\frac{\hbar^2}{2m^*}(\vec{k}_x^2 + \vec{k}_y^2) + \Phi^{EX}{}_{xy}(V,I)\right]$ of all these 2DEG electrons remain identical. Ordinarily, each 2DEG electrons can move in the x-y plane with widely different $\vec{k}_x$ and $\vec{k}_y$. Therefore, chances of $\vec{k}_x$ and $\vec{k}_y$ of each 2DEG electrons taking part in resonant tunneling to be exactly the same are vanishingly small under usual circumstances. Hence, Eq. 2 suggests that $\vec{k}_x$, $\vec{k}_y$ of these 2DEG electrons must converge to a small range of values for any resonant tunneling to be significant. This certainly indicates a similar narrowing of available momentum space of associated indirect excitons at those particular bias values, which are perquisite for having an excitonic BEC. This is because, excitons can be described in terms of two-particle [1] quantum states having excitonic wave vectors $\vec{K}_{Exciton}(V_0) = \vec{k}_e^{2DEG}(V_0) + \vec{k}_h^{InAs}(V_0)$, where $\vec{k}_h^{InAs}(V_0)$ is the quantized momentum of hole in ground state of InAs QD. This $\vec{k}_h^{InAs}(V_0)$ is not a good quantum number for strongly quantum confined QDs. In spite of that, spread of two-particle excitonic momentum $\left(\vec{K}_{Exciton}(V_0)\right)$ must also reduce with narrowing of the spread of $\vec{k}_e^{2DEG}(V_0)$ if substantial resonant tunneling has to occur. Subsequently $\langle \vec{P} \rangle$ of excitons orients towards the $\hat{z}$ direction periodically with increasing voltage bias. In Figs. 4a, 4c, we observe this as increase in photocapacitance oscillation magnitudes with decreasing temperature below ~ 100 K. This is captured in Fig. 4d which illustrates the schematics of $\langle \vec{P} \rangle$ of excitonic dipoles whenever these collectively align along a particular direction at some specific bias voltages during the BEC phase transition. In addition, Fig. 4d also matches with our understandings from the section 4 on coulomb correlated, periodic organization of excitons with increasing bias voltage. Presence of such periodic coulomb correlation at certain bias voltages may actually be aiding the formation of this 'itinerant' excitonic BEC and vice versa. Please note that capacitance measurements at any particular frequency only sample a fraction of dipolar excitons participating in the impedance measurements at that frequency. This fraction increases with decreasing frequencies. Therefore, the fraction of excitons undergoing this BEC phase transition is not calculated from the results obtained at one particular frequency namely 10 kHz.

However, we have also discussed in the last two sections that this excitonic system is periodically driven towards and out of resonant tunneling and BEC state, as more and more charge carriers accumulate near the heterojunction and subsequently tunnel out of it with increasing bias (Figs. 1c, 1d). Any slight variation of energies due to size distribution effects in QDs and in QWs can be compensated by this local variations of $\Phi^{EX}{}_{xy}(V,I)$ in the x-y plane. As long as these small local variations do not affect the final BEC ground state, we expect to see long range phase coherence among these excitons spread over a macroscopically large area. This is certainly observed in these photocapacitance oscillations. Consequently, only those 2DEG electrons with compatible $\vec{k}_e^{2DEG}(V_0)$ can form 0D-2D indirect excitons and can simultaneously take part in coherent resonant tunneling with a constant phase [17] $\theta = \vec{k}_e^{2DEG}(V_0).\vec{z}_{InAs}$, assuming nearly monodisperse InAs QDs. Otherwise, any random phase differences during sequential, incoherent tunneling through millions of InAs QDs could have averaged out these phase coherent oscillations measured over ~200 μm wide area and observed oscillations would have completely washed out. These correlated 2DEG excitons and associated resonant tunneling are, therefore, the main triggers for having such 'itinerant' excitonic BEC in this 0D-2D heterostructures as a function of applied bias. This is a result of direct competition between the single particle momentum uncertainty of localized states of holes with the long range order of the two-particle state of correlated excitons, which eventually dominates over the former.

Spatial anchoring of these indirect excitons with InAs QDs are also crucial to prevent excitonic Mott transitions in this 0D-2D heterostructure. In addition, these InAs QDs act as localized trapping potentials for excitons to sustain this BEC. Estimated dipolar density $\sigma_{ph}$ of 0D-2D indirect excitons as BEC order parameter was calculated in the last section. At first, this value of ~$10^{11}$/cm$^2$ matches well with the measured areal density of InAs QDs in our sample. Moreover, the thermal de-Broglie wavelength ($\lambda_{Th}$) of these excitons can be around several tens of nm at 10 K. As a result, $\lambda_{Th}$ of these indirect excitons easily exceeds the average ~11 ± 2 nm spacing between neighboring InAs QDs with surface density ~$10^{11}$/cm$^2$. It is expected that thermal degrees of freedom of these 0D-2D indirect excitons are mostly associated with lighter electrons in 2DEG within GaAs. So this $\lambda_{Th}$ is approximately estimated using $m_{exciton} \sim m^*$ for localized excitons assuming that heavier holes are confined within InAs QDs under reverse bias. Assuming, effective mass $m^* \sim 0.058 m_0$ in GaAs, where $m_0$ is the free electron mass, we find $\lambda_{Th} = \left(\frac{2\pi\hbar^2}{m_{exciton} k_B T}\right)^{1/2} \approx 100$ nm for T = 10 K, where $k_B$ is Boltzmann constant. However, even with $m_{Exciton} = m_e + m_h$, where $m_e, m_h$ are effective mass of electrons and holes respectively, we find $\lambda_{Th} \approx$ several tens of nm. So it is likely that actual BEC coherence length can be larger than the average spacing between the InAs QDs. As a result, it is expected that wave functions of these quantum coupled, 0D-2D indirect excitons, in fact, begin to overlap below 100 K to facilitate the BEC of excitons. This is also shown in Fig. 4d.

Moreover, it is known that parallel configurations of excitonic dipoles usually support stable dipolar BECs [30]. Repulsive dipolar interactions between excitons mediated by coulomb correlated 2DEG electrons also reduce [31] long-range density fluctuations. Therefore, we predict that this BEC state of dipolar, indirect excitons can be further stabilized [30] at even higher temperatures ($T_0$) with 0D-2D samples having higher in-plane density of QDs and/or using materials having large excitonic binding energies ($E_b$) such that $E_b \gg k_B T_0$. Moreover, it is also known [32] that critical temperature for any two-component excitonic BEC tends to be higher than that of any one-component system due decrease in the net reduced mass over the one component system. This two-component aspect of excitonic BEC will be discussed in section 9. Also see Supplementary Material for further analyses on why these dipolar excitons are expected to undergo

BEC. Consequently, we expect this interacting Bose gas of dipolar, indirect excitons to exhibit quasi long-range order over a macroscopically large area, which will be discussed in the next section.

**7. Off-Diagonal Long Range Order and Interference of Excitonic Polarization Waves**

Correlated photo response over a large, ~ 200 μm wide spatial region can further demonstrate the presence of off-diagonal-long-range-order (ODLRO) which is expected to be associated with such a BEC of excitons. A scheme for phase coherent interference between two macroscopically large, spatially separated groups excitons using spatially fragmented light spots is shown in Fig. 5a. This is qualitatively similar to what is understood in case of atomic BECs [33,34] on how quantum entanglement survives even between spatially dispersed atomic clouds. Here we observe gradual formation and decay of quantum interference 'beats' of photocapacitance as a function of applied bias in Fig. 5b. These beats are formed whenever macroscopically large ensembles of spatially fragmented excitons coherently diffract across the opaque electrical contact with a different in-plane Coulomb correlation energy $\Phi^{EX}{}_{xy}(V, I)$ and consequently oscillate with slightly different frequencies as a function of applied bias. Difference in $\Phi^{EX}{}_{xy}(V, I)$ results in somewhat different 'stiffness' (or the equivalent 'spring constant') for two macroscopically large groups of excitons undergoing interference oscillations. In Fig. 5c, effective spot size of the laser beam was increased slowly beyond the ring periphery by moving a lens to see formations of quantum interference beats in photocapacitance. The contrast $\left[\sim \frac{(C_{Max}- C_{Min})}{(C_{Max}+C_{Min})}\right]$ of photocapacitance based interference oscillations as a measure of first order spatial correlation of exciton numbers measured over a macroscopically large area actually increases (Fig. 5d) with increasing exciton density (or light intensity used for photoexcitations). This indicates a density driven enhancement of this contrast. We knew [14,15] that photocapacitance as $\langle \vec{P} \rangle$ can probe the number density of excitonic dipoles. Hence, enhanced contrast of these interference oscillations are related to increasing 1st order correlation of number densities of excitons or 2nd order correlation $g^{(2)}$ of excitonic matter wave amplitudes. Use of incoherent white light also indicates that such phase coherent interferences are not originated from any polarization interference of the light used in photoexcitation either.

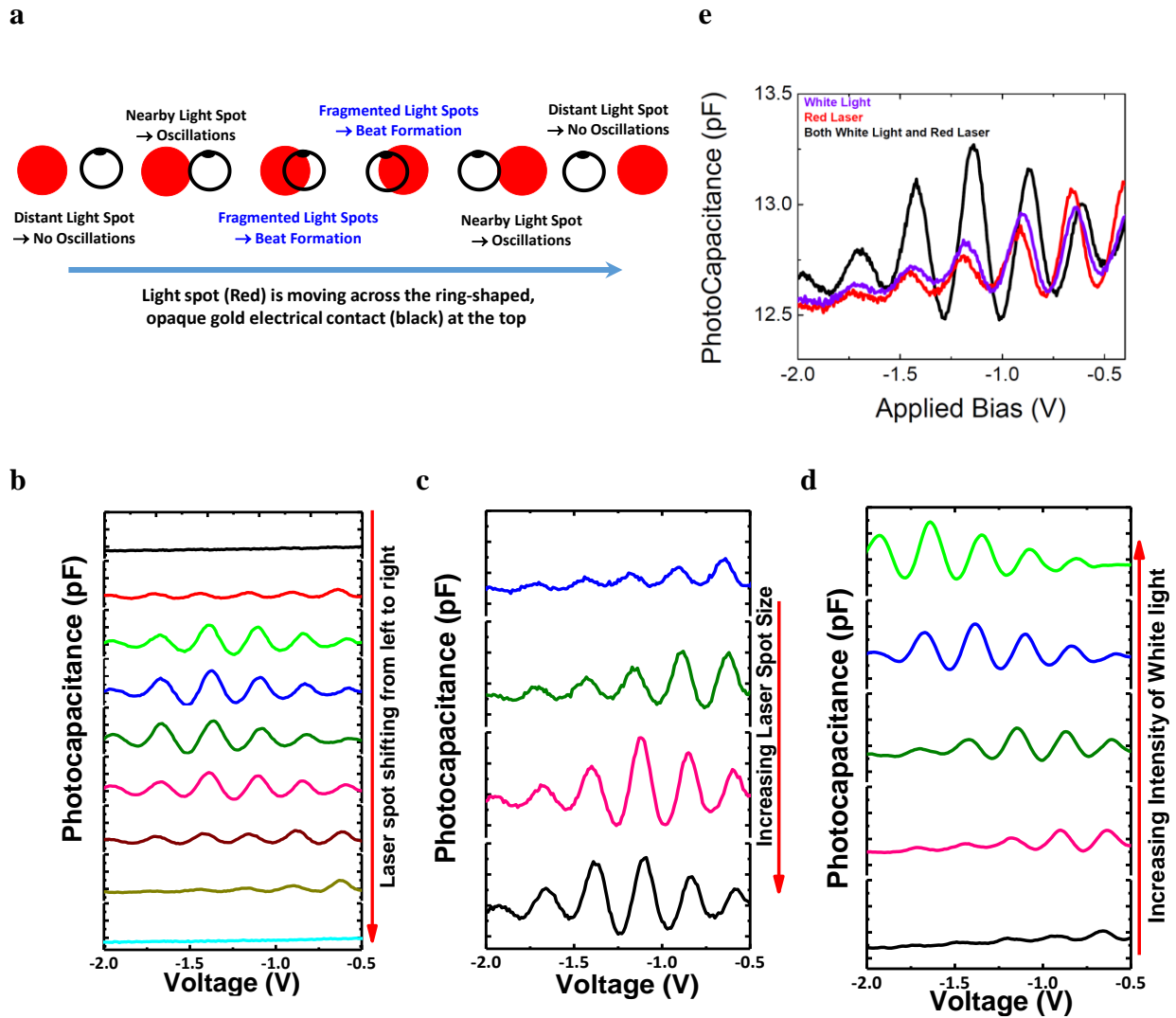

**Fig. 5.** Interference [33,34] of excitonic matter waves and evidence of long range spatial correlation. (a) Schematic representation shows how the 633 nm, red laser spot is moving from left to right across the ring-shaped top electrical contact of diameter ~200 μm. Spatially fragmented light spots (red) create two separate groups of excitons which then interfere across the vicinity of gold ring (black) of width ~25 μm. (b) Formation and decay of quantum beats in photocapacitance whenever spatially separated excitonic ensembles are interfered. (c) Spot size of the laser beam is gradually increased across the opaque ring to fragment the excitons in two groups and we consequently see formation of quantum interference beats among these in photocapacitance. (d) Observation of quantum interference beats even with incoherent white light rules out any polarization interference of the light itself as the cause. (e) Most importantly, quantum interference beats with high contrast is also seen when two separate, uncorrelated light sources (633 nm and a white light source) are used to photo excite the sample simultaneously.

Subsequently, in Fig. 5e, we establish quantum beats of interference oscillations with high contrast when a continuous wave (CW) laser beam of 633 nm and a purely white light source are used simultaneously. Intensities of these light sources were adjusted to get similar photocurrent values. Oscillatory changes in photocapacitance beats at simultaneous photo excitations are clearly much larger than the simple algebraic sum of photocapacitance produced by individual light beams once the dark capacitance background is subtracted from the photocapcitance traces. This phase coherent enhancements indicate some form of 'interference' effects during simultaneous photo excitations. As such, interference of two independent, pulsed lasers can happen only within times scales smaller than their coherence times. Therefore, these observations of quantum beats of interference oscillations even with incoherent white light (Figs. 5d, 5e) convincingly prove that spatially long range phase 'coherence' witnessed by these photocapacitance based interference oscillations are not only long lived, but also not originating from coherence of the photoexcitation source itself. In a way, observation of such spatially long range phase 'coherence' is most likely due to formation of coherent, excitonic matter wave states. We consider this robust phase correlation over such a macroscopically large area as a direct measure of the ODLRO [33,34] of BEC of excitons. However, such density dependent beat patterns are fully absent in photo-G/ω (see Fig. S2 in Supplementary Material). This is because photo-G/ω tends to follow bulk $dI_{Ph}/dV$ sequentially as shown in Fig. 1d, $I_{Ph}$ being the DC photocurrent which is generated everywhere within the sample structure and not necessarily originated from excitons formed around the 0D-2D heterojunction. Therefore, photocapacitance is advantageous in detecting such spatially long-range, cooperative phenomena involving $\langle \vec{P} \rangle$ of excitonic dipoles. We will further elaborate on this unique characteristic of photocapacitance in section 11.

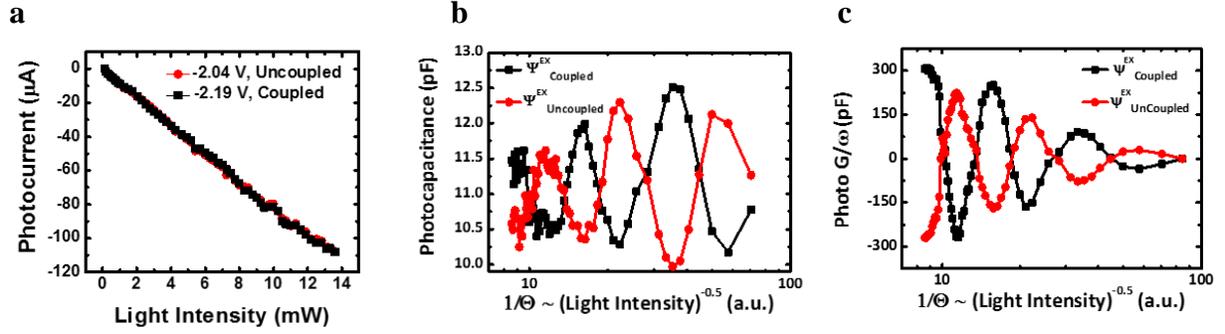

**Fig. 6.** Rabi oscillations of two-component BEC of excitons. (a) Photocurrent does not oscillate with increasing light intensity (I) at all. This indicates that applied voltage bias strongly pins the energy levels and subsequently and there is no change in the resonant tunneling condition with increasing light intensity. However, (b) Photocapacitance and (c) photo-G/ω as functions of $1/\Theta \sim 1/\sqrt{I}$ oscillates coherently as in [35-39]. $1/\Theta$ is also proportional to ~time. Such Rabi oscillations of these two-level macroscopic quantum states of $\psi^{EX}_{Coupled}$ and $\psi^{EX}_{Uncoupled}$ are out of phase with each other in both photocapacitance and photo-G/ω plots.

## 8. Experimental signature of Rabi Oscillation of Two-component Excitonic BEC

Excitonic Rabi oscillations as decaying oscillations in photoresponse versus photoexcitation amplitude were mostly reported [35-39] using 'single' QDs only. Here we will present experimental results on Rabi oscillations of photocapacitance as a function of photoexcitation intensity measured over millions of QDs as additional evidence of the presence of a macroscopically large quantum state of excitonic BEC. We know that exciton densities in our sample are changed with both increasing bias voltages and with increasing light intensities. Till now, we have only reported oscillations of photocapacitance with respect to applied bias in previous figures. Fig. 6a shows that dc photocurrent increases monotonically with increasing light intensity at -2.04 V (-2.19 V) corresponding to one maximum (minimum) of the photocapacitance vs bias oscillations. This is because of the strong pinning of energy levels of the 0D-2D structure with applied bias voltage. However, in Figs. 6b and 6c, we display photocapacitance and photo-G/ω oscillations measured at -2.04 V and -2.19 V as functions of $(I)^{-0.5}$, where $I = |\vec{E}|^2$ and $|\vec{E}|$ is electric field amplitude of 'continuous' photoexcitations using a 633 nm He-Ne laser. These are clearly 180° out of phase with each other. Therefore, this observed phase coherent oscillations of photocapacitance as a function of photoexcitation intensity measured over millions of QDs reinforces our argument that these are certainly not just any simple-minded sequential tunneling whenever the quantized levels of InAs QDs and GaAs 2DEG come into and out of resonance with each other as prescribed [8-10] earlier. It was also known [40-42] that even with 'continuous' excitations, steady state Rabi oscillations of two-level quantum systems can persist. Therefore, we interpret the observations of Fig. 6b as collective Rabi oscillations of *'macroscopically'* large, quantum states of excitonic BEC. Detailed description of this two level quantum system involving $\psi^{EX}_{Coupled}$ and $\psi^{EX}_{Uncoupled}$ and bias induced quantum coupling and uncoupling of macroscopic quantum states of 0D and 2D structure will be presented in Section 9 as further supporting evidences of the 'quantum' nature of these excitonic phenomena probed with photocapacitance. We will present photocapacitance based optical spectra in the next section to show that the line widths of two excitonic peaks presented in Fig. 7a are, although somewhat wider, but certainly small compared to splitting of peaks by 100 meV to enable the observation of these Rabi oscillations.

As such, the Rabi frequency is $\Omega_R = |\vec{\mu}_{12}.\vec{E}/\hbar|$ and pulse-area is $\Theta = \left|\vec{\mu}_{12}.\left[\int_0^{\Delta\tau} \vec{E}(t)\left(\frac{dt}{\hbar}\right)\right]\right| = |\vec{\mu}_{12}.\vec{F}|$, where $\vec{\mu}_{12} \cong \vec{\mu}_{12}(V,I)$ is the electric dipole matrix for excitonic transition, $\Delta\tau$ is the duration probed by our measurements. Clearly, $\frac{1}{\Omega_R}$ is inversely proportional to the square root of light intensity, which is also proportional to time. Usually Rabi oscillations are seen with pulsed lasers at times shorter than dephasing time $(\tau_\gamma \sim 10^{-12}\ s)$. So Fig. 6b illustrates negligible [43,44] dephasing rate $\left(\frac{1}{\tau_\gamma}\right)$ of this excitonic system such that $\Theta \gg \tau_\gamma^{-1}$. Therefore, observed Rabi oscillations measured with photocapacitance actually indicates 'insignificant dephasing' even at this temperature and time scales probed in our steady state photocapacitance measurements. This fact certainly endorses the presence of excitonic BEC with its characteristically lesser dissipations to allow these oscillation as a function of increasing light intensities. Any normal phase transition from an uncorrelated excitonic phase to an ordered ferroelectric like transition without any quantum coherence need not account for these observations.

## 9. Periodic coupling and uncoupling of 0D-2D Quantum Structure via Tunneling & Hadamard gate operation.

It was shown earlier (Fig. 2c) that we hardly observe any excitonic peaks under zero bias at 10 K for the optical spectra measured with photocapacitance and photo-G/ω. However, in Figs. 7a and 7b, we show spectra at two successive bias values of maximum (-0.40 V) and minimum (-0.53 V) of photocapacitance or $(\vec{P})$ oscillations (from Fig. 1c). Presence of two resonant peaks in photocapacitance spectra of Fig. 7a clearly indicates the role of bias driven formation of excitons. Whereas, in Fig. 7b, we see a somewhat broader photocapacitance peak in the middle. On the other hand, both of these excitonic peaks survive in the photo-G/ω spectra at both of these biases (also see Fig. S3 in Supplementary Material). Further comparisons with Fig. 2c show that these excitonic transitions centered at ~1.53 eV and at ~1.61 eV are not predominantly originated from any optical generation of direct excitons within InAs QDs and also in GaAs layers in the absence of any applied bias.

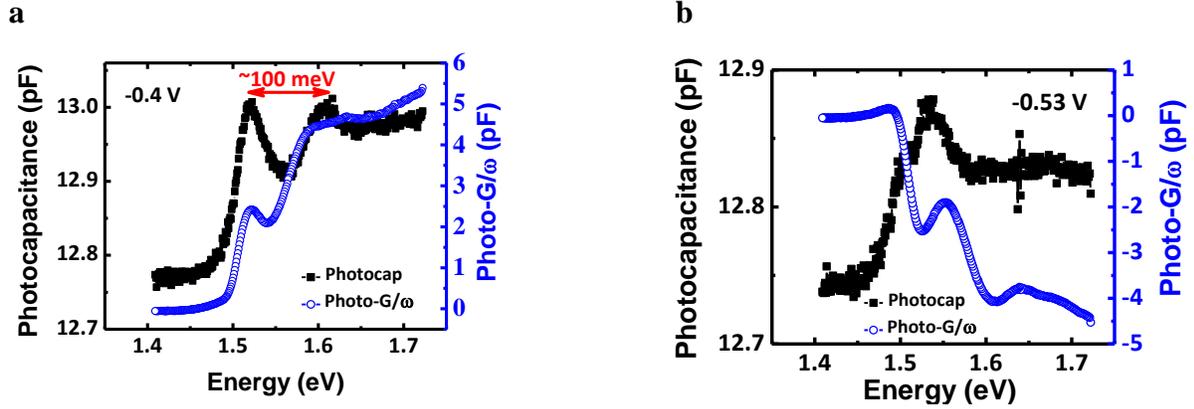

**Fig. 7.** Quantum coupling (decoupling) of 0D-2D heterostructure. (a) Photocapacitance and photo-G/ω spectra under -0.40 V bias corresponding to maximum of photocapacitance oscillation (Fig. 1c). We observe the presence of two sharp excitonic peaks (black) around 1.53 eV and 1.61 eV in both spectra. (b) Photocapacitance and photo-G/ω spectra under -0.53 V bias corresponding to a minimum of photocapacitance oscillation (Fig. 1c). Unlike the photo-G/ω, those two distinct excitonic peaks of Fig. 7a are strikingly missing only from the photocapacitance spectra in Fig. 7b. Similar results (Fig. S3 in Supplementary Material) are witnessed for all other applied biases corresponding to maximum or minimum of photocapacitance oscillations from Fig. 1c.

We observe the doubly split excitonic spectra only at specific bias voltages whenever the capacitive coupling (i.e. $\langle \vec{P} \rangle$) is highest at maximum photocapacitance in the necessary presence of exciton BEC. Every time excitons either collectively assemble (or disassemble) into (from) a Bose condensate state around the QDs in presence of excess carriers with changing biases, these produce corresponding changes in photo current $I_{Ph}$. So, spectral signatures of optical generation of indirect excitons survive in the photo-G/ω (~$dI_{Ph}/dV$) spectra at both biases in Figs. 7a and 7b.

Moreover, photo generated excitons elsewhere (in bulk GaAs) in the sample also contribute to photo-G/ω. As a result, photo-G/ω spectra (Fig. 7) doesn't necessarily reflect collective changes of dipolar excitons accumulated at the 0D-2D heterojunction. However, such periodic presence and absence of excitonic peak splitting in photocapacitance spectra as a function of applied bias is certainly very intriguing to say the least.

Next, we will explain why periodic changes in the photocapacitance spectra also indicate quantum coupling-decoupling of the 0D and 2D heterostructure and how all these can be modeled as a Hadamard quantum Gate operation. Due to their smaller effective mass, electrons take part in quantum tunneling more efficiently than holes. So we assume that the usual Hadamard Operator $\widehat{H}(V,I) = \frac{1}{\sqrt{2}}\begin{bmatrix} 1 & 1 \\ 1 & -1 \end{bmatrix}$ acts only on electron wave functions ($\varphi_L^e$ and $\varphi_R^e$) and not on hole wave functions ($\varphi_L^h$ and $\varphi_R^h$) of these excitons, where subscripts $L \rightarrow$ InAs QD and $R \rightarrow$ GaAs 2DEG under reverse bias. Here $V, I$ are applied bias voltage and photo excitation intensity respectively. If excitonic wave function is

$$\varphi^{Ex}(r_e, r_h) = \frac{1}{\sqrt{N}} \sum_{k_e, k_h} a(k_e, k_h) \varphi_{k_e}(r_e) \varphi_{k_h}(r_h) \qquad (3)$$

where $a(k_e, k_h)$ is the probability amplitude. Then relevant excitons can be: (a) direct excitons inside InAs QD as $\varphi^{DX} = \varphi_L^e \varphi_L^h$ and (b) indirect excitons with holes inside InAs QD and electrons inside GaAs 2DEG as $\varphi^{IX} = \varphi_R^e \varphi_L^h$ under the reverse bias. Here the summations over $k_e$ and $k_h$ in $\varphi^{Ex}(r_e, r_h)$ drop out during BEC as $|\vec{k}_e^{2DEG}(V_0, I_0)| \rightarrow |\vec{k}_h^{InAs}(V_0, I_0)| = k_0\ (constant)$. Ground state energy of 2DEG and consequently that of indirect excitons (IX) vary with applied bias. Specifically, in such a double well potential [45], quantum coupling of 2DEG and the QDs at bias values of maximum photocapacitance oscillation can be written as an operation of the Hadamard gate on the uncoupled quantum state as

$$\widehat{H}\boldsymbol{\psi}^{EX}_{Uncoupled} = \widehat{H}\begin{pmatrix}\varphi_{DX}\\ \varphi_{IX}\end{pmatrix} = \begin{pmatrix}\varphi_+^{EX}\\ \varphi_-^{EX}\end{pmatrix} = \boldsymbol{\psi}^{EX}_{Coupled} \qquad (4)$$

where $\varphi_+^{EX} = \varphi_+^e \varphi_L^h = \frac{1}{\sqrt{2}}[\varphi_L^e + \varphi_R^e]\varphi_L^h$ and $\varphi_-^{EX} = \varphi_-^e \varphi_L^h = \frac{1}{\sqrt{2}}[\varphi_L^e - \varphi_R^e]\varphi_L^h$ are the entangled states of 0D-2D indirect excitons under maximum coupling. Correspondingly, $\boldsymbol{\psi}^{EX}_{Coupled}$ is the ground state of the two-component BEC at bias voltages of maximum photocapacitance. If these quantum coupled states have energies $E_+$, $E_-$, then experimentally we can recognize this energy splitting $|E_+ - E_-|$ as text book like splitting of degenerate energy levels of any double quantum well [45]. Amount of this energy splitting depends on the strength of coupling between the 0D and 2D quantum reservoirs. Consequently, it is tempting to identify this with doubly-split excitonic peaks with $|E_+ - E_-| \approx 100\ meV$ in the photocapacitance spectra shown in Fig. 7a (also see Figs. S3a, S3b of Supplementary Material). This $\widehat{H}(V,I)$ acts back on two component BEC state of $\boldsymbol{\psi}^{EX}_{Coupled}$ to yield $\boldsymbol{\psi}^{EX}_{Uncoupled}$ at biases of minimum photocapacitance coupling and the doubly split excitonic peaks vanishes from the photocapacitance spectra leaving aside a broad peak ~1.55 eV (Fig. 7b and Figs. S3c, S3d of Supplementary Material). This broad composite peak originates at bias value of minimum photocapacitance couplings because photo generations in p and n type GaAs layers, in GaAs 2DEG and also in InAs QDs are also contributing to formation of excitons around the AlAs potential barrier.

As mentioned above, the line widths of these exciton peaks are not narrow enough for a typical BEC. Sharpness of the measured photocapacitance spectra are currently limited by the intensity of our white light source which required us to keep the monochromator slits open to the fullest extent ($\Delta\lambda \sim 9$ nm). On the other hand, we see much sharper excitonic spectral features in photoluminescence measurements using strong lasers under open circuit conditions to ensure the quality of our sample and formation of energetically sharp excitonic levels. However, these photoluminescence spectra and their own oscillations with respect to applied bias are not being presented here. These results will be submitted as part of a separate manuscript in future. In addition, we are also not denying any possible influences of tunneling induced depletion of excitons from the BEC state as another important factor for observing such wider excitonic peaks in these photocapacitance spectra. Moreover, in this 0D-2D sample, apart from 'itinerant' exciton BEC at specific bias values, we also have the presence of – (a) coherent resonant tunneling, (b) coulomb correlated excitons and (c) coupling and uncoupling of 0D QDS and 2DEG in such quantum coupled double-well like structure. The third effect, specifically, delocalizes the electrons in the reverse bias within the 0D and 2D wells and can also cause broadening. In fact, towards the end of this manuscript, we will argue why quantum coherence sustains even in presence of tunneling in Section 12. Therefore, here we understood these spectral evolutions in terms of excitonic peak splitting at the bias values of maxima of photocapacitance oscillation. This actually corresponds to quantum coupling and subsequent splitting of ground state energy of two-component excitonic BEC in this 0D-2D heterostructure. Moreover, phase coherent photocapacitance oscillations slowly decay (Fig. 1c) at increasing biases in presence of enhanced numbers of accumulated excitons at the AlAs junction and we also see broader peaks (Figs. S3a, S3b in Supplementary Material) with increasing biases.

**10. Experimentally Tunable Bloch Sphere of Excitonic BEC State**

In Figs. 8a, 8b, we see that both photocapacitance and photo-G/ω respectively oscillate as a function of photo excitation intensity. At first, this is similar to Rabi oscillations of the coupled and uncoupled state at maximum (-2.04 V), minimum (-2.19 V) of collective electrical polarization of excitons reported above (Fig. 6). In fact, these quantum oscillations of ac photo responses are observed even when measured dc-photocurrent hardly oscillate with increasing light intensity (see Fig. S4, Supplementary Material).

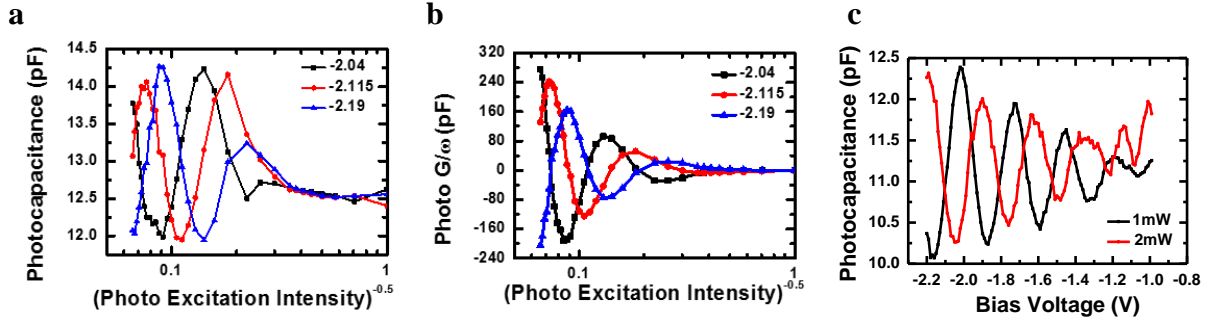

**Fig. 8.** Experimentally tunable Bloch sphere of excitonic BEC. Phase coherent Rabi oscillations (as in [35-39]) measured with – (a) photocapacitance and (b) photo-G/ω as a function of inverse of the square root of the photo excitation intensity in arbitrary units at applied bias voltages corresponding to maximum (-2.04 V), minimum (-2.19 V) and in-between (-2.115 V) biases at one particular light intensity. The coupled (black) and uncoupled (blue) state respectively at biases of maximum and minimum photocapacitance are certainly out of phase with each other. The trace (red) of phase coherent evolution of photo response at an in-between (-2.115 V) bias voltage is also in the middle. (c) Out of phase evolution of two orthogonal quantum states of excitonic polarization as a function of bias voltage can also be created using two different light intensities.

Periodic variations in coherent 'resonant' tunneling in 0D-2D heterostructure necessarily needs 'itinerant' excitonic BEC as argued above. However, resonant tunneling can certainly not happen at all and in-between bias voltages as alignment of 0D-2D discreet energy levels crucially depend on that. However, photo generated and bias driven excitons can still form at the 0D-2D heterojunction at any finite biases and also undergo in-plane correlation effects based on $\Phi^{EX}_{xy}(V, I)$ and subsequent formation of 'itinerant' excitonic BEC. Nonetheless, we still see thesequantum oscillations at -2.115 V as a function of increasing exciton density with increasing light intensities as well. This also strongly indicates that the underlying cooperative phase of $\langle \vec{P} \rangle$ of these dipolar excitons and the associated two-level quantum state of excitonic BEC as described in sections 8,9 can be tuned with both applied bias and photoexcitation intensity. Moreover, it allows us the flexibility of defining the orthogonal basis states of this two-level quantum state of excitonic BEC with applied biases as shown in Figs. 8a, 8b. Additionally, one can also create phase coherent and nearly orthogonal (out of phase) quantum states of $\langle\vec{P}\rangle$ of excitonic BEC at two different photoexcitation intensities and probe the same using photocapacitance as a function of bias voltage. This is demonstrated in Fig. 8c. Such operational freedom to experimentally define the orthonormal basis states of the Bloch sphere of excitonic qubits was never reported previously.

**11. Why photocapacitance can measure the Multi-partite nature of Excitonc BEC state?**

So far this 0D-2D heterostructure had helped us to identify the presence of excitonic BEC coexisting with resonant tunneling. Now we demonstrate why photocapacitance [14,15] is so unique in its ability to probe Schrödinger's Cat like macroscopically large, multi-partite quantum state of $\vec{P}$ of excitonic BEC. Finally, we will propose how all these can be used to build abundantly large quantum register of excitonic qubits.

It is not always advisable to visualize a physical situation in terms of any classical picture of excitons as localized dipole of electron-hole pairs, specifically when we are dealing with quantum phenomena like excitonic BEC, quantum coupling of QDs and 2DEG, coherent tunneling and quantum capacitance etc. So we now try to understand why photocapacitance can probe such macroscopic quantum state of excitonic BEC. Oscillatory changes in photocapacitance ($C$) actually represent variations of quantum mechanical average $(\langle\hat{P}\rangle)$ of the vector operator for electric polarization $\widehat{P}$ of these dipolar excitons in the quantum ground state of BEC. One can write this as

$$\langle \widehat{P} \rangle = Tr[\widehat{P} \rho(t)] = Tr[\widehat{P} \exp(-i\hat{h}t/\hbar)\exp(i\vec{F}.\widehat{P})\rho(t_0)\exp(-i\vec{F}.\widehat{P})\exp(+i\hat{h}t/\hbar)] \quad (5)$$

where [46] ρ(t) is the density matrix of the BEC state and $\hat{h}$ is the interaction Hamiltonian under rotating-wave approximation. If the dipole moment operator of j$^{th}$ exciton in BEC quantum state is $\hat{p}_j \sim (\vec{\mu}_{12})_j$ and $\widehat{P} = \sum_j \hat{p}_j$, then

$$\exp(-i\vec{F}.\widehat{P}) = \exp(-i\vec{F}.\hat{p}_1)\exp(-i\vec{F}.\hat{p}_2) \ldots \exp(-i\vec{F}.\hat{p}_n) \quad (6)$$

As a result, the calculation of $\langle \widehat{P} \rangle$ is multipartite in nature and factorable in terms of 'n' quantum mechanical operators. Therefore, photocapacitance oscillations, in principle, can detect the n-particle quantum correlations necessary [47] to establish the macroscopically large 'Schrödinger's Cat' like quantum state of excitonic BEC (See sections 2,4 of Ref. 47). In fact, we have already demonstrated (in Fig. 5) diffraction/interference of spatially fragmented but macroscopically large population of excitonic matter (polarization) waves, which is necessary for [33,34,47] establishing the presence of Schrodinger cat states of excitons.

**12. Why Macroscopic Quantum Coherence of Excitonic BEC state survives even in presence of Tunneling induced Decoherence?**

In practice, there are many physical situations in which macroscopic systems show quantum tunneling without quantum coherence. Incoherent tunneling of electrons under reverse bias can actually disturb the excitonic BEC which is a 'necessary' condition for any coherent resonant tunneling in such 0D-2D structure as explained above using Eq. 2 (Ref. 29, Fig.1 ibid.).

However, it was known [47] that any observation of quantum coherence & tunneling between macroscopically different states can also indicate Schrödinger's Cat like quantum states. Main difficulties in achieving that comes from dissipative coupling with the environment which eventually lead to strong decoherence. Nevertheless, if we are dealing with a quantum state with very small dissipation, possibly a BEC state then it can preserve that coherence. Under reverse bias, the electrons can move back and forth between the 0D and 2D potential wells with frequency ~2$\Gamma$. Equivalently, this leads to the splitting of the BEC ground state as a doublet [45] with energy difference $\Delta$ ~2$\hbar\Gamma$ as also mentioned above (see Fig. 7a). In thermal equilibrium conditions, standard tunneling probability of the resonance oscillation is ~exp(-$\gamma$t). The coherence of tunneling process can be lost when the damping $\gamma = \omega_0^{-1}\Gamma^2$ multiplied by $\frac{k_B T}{\hbar \Gamma}$ (if this is >> 1) becomes comparable to $\frac{\Gamma}{2\pi}$ [47] (See section 5 of Ref. 47 for more details). This is because, time-scales for these two phenomena can be quite different. Experimental control of quantum coherence requires that the relative phase of the wave function should be preserved over times of order $\Gamma^{-1}$. In principle, this $\Gamma^{-1}$ can be very long for a BEC state. Any 'observation' of the excitonic system at time intervals shorter than $\Gamma^{-1}$ can destroy these quantum coherence effects and effectively localize the system on one side of the barrier in the double quantum well system. In other words, one can say that [47] quantum coherence will be totally lost if the root mean-square fluctuations (at frequency $\Gamma$) of the difference in energy between the two originally degenerate states after splitting becomes comparable to the tunneling energy $\hbar\Gamma$. Assuming, $\Delta$ ~ 100 meV from Fig. 7a, $\omega_0$ ~ 1.5 $eV/\hbar$, T=10 K we get $\gamma$ ~$10^{12}$ Hz and $\Gamma$~ $10^{14}$ Hz. We clearly see that $k_B T \ll 2\hbar\Gamma$ and $\frac{k_B T}{\hbar \Gamma}$ ~ 1/100. As a result, $\left(\gamma \frac{k_B T}{\hbar \Gamma}\right)$ ~$10^{10}$ Hz and $\frac{\Gamma}{2\pi}$ ~ $\frac{10^{14}}{2\pi}$ Hz. As a result, $\left(\gamma \frac{k_B T}{\hbar \Gamma}\right)$ ~$10^{10}$ Hz is certainly $\ll \frac{\Gamma}{2\pi}$ ~ $\frac{10^{14}}{2\pi}$ Hz. Therefore, these quantitative estimates clearly support [47] why macroscopic quantum coherence of the two-component exciton BEC state is sustained in this quantum coupled 0D-2D heterostructure even in presence of tunneling.

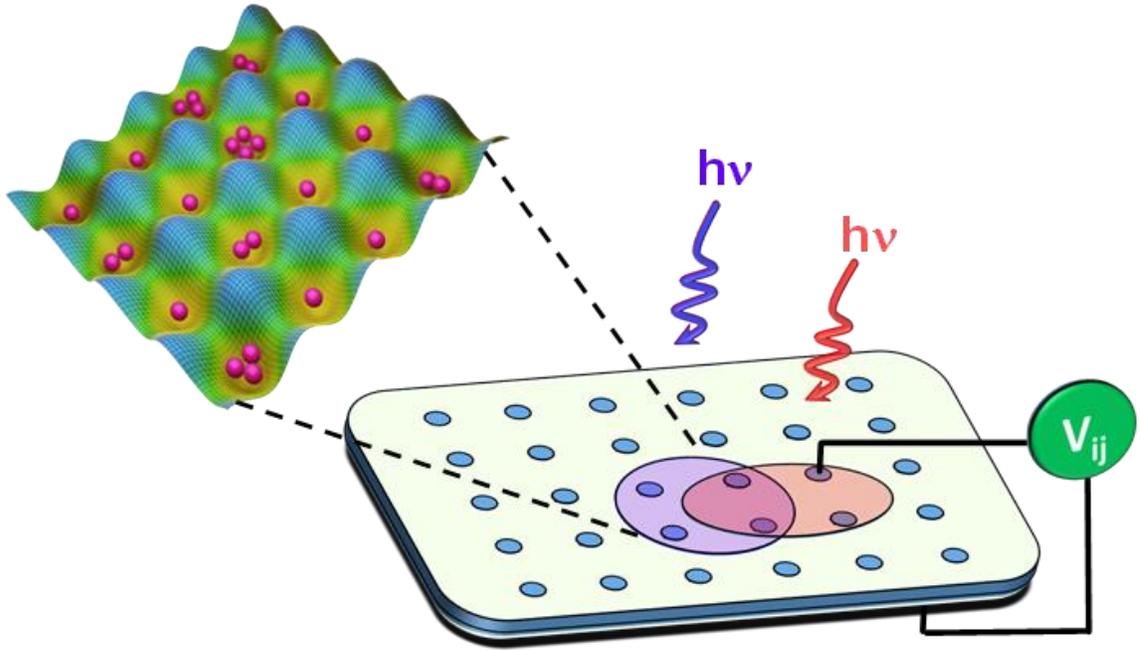

**Fig. 9.** Diagrammatic representation of a quantum register. A representative sample with an array of circular, semitransparent electrical contact pads (blue) on the top surface of this 0D-2D heterostructure (as in Fig. 1a). Each pad in the array (i,j) can be used to apply voltage bias ($V_{ij}$) as well as to monitor photocapacitance signal. Light spots (purple or orange) are being used to generate 'itinerant' excitonic BEC as a function of applied bias. Formation of BEC of excitons (pink spheres) as $(\psi_{Coupled}^{EX})^{n_1} \otimes (\psi_{Coupled}^{EX})^{n_2} \otimes \ldots \ldots (\psi_{Coupled}^{EX})^{n_N}$ using a 0D-2D heterostructure within a light spot is schematically shown in the top left corner. Dashed black lines are just a guide to eyes towards the region within the (purple) light spot below which such excitonic BEC is formed around that 0D-2D heterojunction. The bias voltage ($V_{ij}$) dependent potential energy landscape is generated due to coulomb correlation energy and formation of two-component BEC of dipolar excitons during coherent resonant tunneling through this 0D-2D heterostructure. Pink colored spheres represent $n_{ij}$ excitons within the 2D 'checkerboard' like potential below the ij[th] contact pad with all $n_{ij} \gg 1$. A N-qubit quantum register having N numbers of electrical contact pads (blue) can be used, where N can easily be $\gg 1$. This N-qubit quantum register can also be configured as $(\psi_{Coupled}^{EX})^{n_1} \oplus (\psi_{Uncoupled}^{EX})^{n_2} \oplus \ldots (\psi_{Coupled}^{EX})^{n_N}$ etc. combinations whenever multiple light spots (purple and orange) overlap. Such overlapping lights spots (purple and orange) can even produce interference [33,34,47] of excitonic matter wave in a quantum coherent way as shown in Fig. 5e. These photons can have same or different energies based on required functionality as shown in Figs. 1e, 1f. As a result, the 0D-2D structure is maximally quantum coupled below some pads and not coupled in the other pads using locally different applied biases ($V_{ij}$) and light intensities. Shapes, sizes and overall design of the array of electrical contact pads and the distance between two neighboring pads can be customized using micro/nano lithography for different quantum device applications.

## 13. How to fabricate a very large quantum register with 'itinerant' BEC of millions of excitons?

Experimental detection of macroscopically large, quantum coherent state within semiconductor heterostructures will open up new paradigms to study emergent quantum phenomena in condensed matter physics and to explore the basic physics of quantum entanglements using semiconductor structures. Usually, bottom-up approaches are used in quantum computation, where one assembles individual qubits into a quantum superposition state to fabricate N-qubit quantum registers in a step-by-step manner. However, here we explored a top-down approach where macroscopically large, quantum coherent state of excitonic Bose-Einstein condensate (BEC) having millions or more dipolar excitons as qubits can be experimentally controlled to form N-qubit quantum registers in the temperature range of 10-100 K. Such systems can offer better computational capabilities and sufficient redundancy [48] during quantum error corrections and is already within the operational temperature ranges of some cryogenically cooled, commercial supercomputers.

Operation of any quantum register is qualitatively based on identifying qubits where a physical system mimics a two-level quantum system. This also requires plausible quantum gate operations (section 9) as well as experimental control of Rabi oscillation of this two level quantum state (sections 8, 10). Experimental results presented in Fig. 7 of section 9 actually highlight how these excitonic matter waves can even be entangled (coupled) as well as disentangled (uncoupled) using applied bias voltages. Therefore, in Fig. 9, we propose an operationally executable schematics for N-qubit system using the two-component BEC [49,50] of excitons within a 0D-2D heterostructure to build a large (N >> 1) quantum register consisting of bosonic clones of many excitons in the quantum ground state of that BEC. It was also argued [49] that such large N-qubit BECs can allow faster quantum gate operations due to bosonic enhancement of energy scales. Such large quantum registers can, in principle, be fabricated with multiple, overlapping light spots and applied biases and probed with localized sensing of $\langle \widehat{P} \rangle$ using photocapacitance. This is possible because one can use applied bias over a confined region to selectively modulate the quantum state of some excitons undergoing BEC independently of the rest. One can also use a combination of overlapping light spots to collectively interfere and entangle excitonic matter waves as shown in Fig. 5e. So the abundant multiplicity (millions or more) of excitonic qubits in macroscopically large, coherent, quantum superposition(s) can allow experimentally tunable quantum entanglement. We also speculate that such plentiful number of excitons (as qubits) residing in a single quantum ground state of excitonic BEC can enhance computational capacity and can, on paper, permit adequate redundancy for efficient, fault-tolerant quantum operations. There were past studies on fault tolerant operations of bosonic qubits [51-53], however none actually used the concept of using bosonic clones in macroscopic quantum state of a BEC explicitly. In principle, most of the DiVincenzo's criteria [54] for universal quantum computation can actually be fulfilled in this 0D-2D system by controlling the local electric fields around the quantum dots and also by varying photoexcitation intensities of overlapping light spots (see Fig. 9). Further work is going on to fabricate a full fledge quantum computer having universal quantum gate operation using such excitonic systems.

## 14. Conclusions

We provided ample corroborative experimental evidences including oscillations in quantum capacitance as a measure of collective, phase coherent, undulations of average electrical polarization $\langle\widehat{P}\rangle$ of excitonic dipoles over a large area. These include spontaneous enhancement of $\langle\widehat{P}\rangle$ below ~100 K, periodic splitting of excitonic photocapacitance spectra as a result of quantum coupling(decoupling) between the 0D-2D structure, interference of excitonic matter waves as proof of ODLRO as well as light intensity dependent Rabi oscillations to establish our proposition on experimental control of a macroscopically large quantum state of excitons. All these experimental results are meticulously analyzed using somewhat uncommon but logically interconnected, consistent framework. Moreover, all of those experimental evidences presented in different sections actually complement each other and also support our understanding of many-body cooperative effects in this 0D-2D excitonic system in the form of an 'itinerant' BEC of excitons coexisting with coherent resonant tunneling.

Thereafter we proposed how such experimental control of macroscopic quantum state of 'itinerant' two-component BEC can be used generate excitons as entangled qubits. We suggested that these excitonic two level states as qubits can be tuned with both applied voltage bias and photo excitation intensity to build very large quantum registers having millions of excitons as bosonic clones. As such, the operational temperatures of these excitonic BEC based quantum registers can be raised further with more densely packed, ordered array of QDs in the x-y plane and/or using materials having larger excitonic binding energies. Most importantly, many recent experimental observations of excitonic BECs are reported/predicted in both 2D materials [55-58] and also in bulk materials [59]. Some of these 2D material systems can certainly be used with overlapping lights spots to produce interference [33,34, 47] oscillations of excitonic matter wave in a quantum coherent way as shown in Figs. 5e. However, 'localized' experimental control of Hadamard like quantum gate operation using macroscopic quantum state of excitonic BEC by tuning the quantum coupling of 0D and 2D structures with applied bias (as in section 9) are likely not viable using only 2D planar materials/structures. It is mostly because charge neutral excitons in these planar 2D structures are particularly difficult to confine in the x-y plane. This is, however, not an issue in this 0D-2D heterostructure. As mentioned in section 6, anchoring of these 0D-2D indirect excitons with InAs QDs are also beneficial to prevent excitonic Mott transitions in this 0D-2D heterostructure. In fact, these InAs QDs provide localized trapping potentials for 0D-2D indirect excitons to facilitate this excitonic BEC as well.

As such, the science and technology of fabrications of single crystals of 0D-2D quantum heterostructures using 2D materials (e.g. transition metal di-chalcogenides, oxides, perovskites etc.) having higher excitonic binding energies are still an open challenge in material science. Therefore, we hope that our experimental results on quantum control of excitonic BEC in 0D-2D hybrid heterostructure should motivate further intensive research works to address fabrication challenges of such 0D-2D structures using materials having higher excitonic binding energies in future.

Even now these 0D-2D heterostructures can be miniaturized as well as scaled up using the existing III-V and/or Nitride based semiconductor fabrication technologies. Such feasibilities of fabricating quantum devices, however, certainly difficult to implement for most other current approaches being developed for quantum computation outside the boundaries of a sophisticated research laboratories. Therefore, our study not only brings the physics and technology of Bose-Einstein condensation within the reaches of semiconductor based optoelectronics chips, but also opens up experimental investigations of the fundamentals of quantum physics using similar techniques.

Thus, we are hopeful that the research work presented in this manuscript can certainly initiate new experimental efforts both in the synthesis/growth and optoelectronic investigations of these new class of 0D-2D quantum materials/heterostructures. Furthermore, it has prospects of generating newer quantum technological applications in building next generation quantum registers using semiconductor chips. In addition, we hope, these experimental results will also act as a genuine trigger for more detailed theoretical efforts to develop rigorous models to understand the above mentioned many body quantum cooperative effects of excitons in 0D-2D quantum materials/heterostructures.

**Materials and methods**

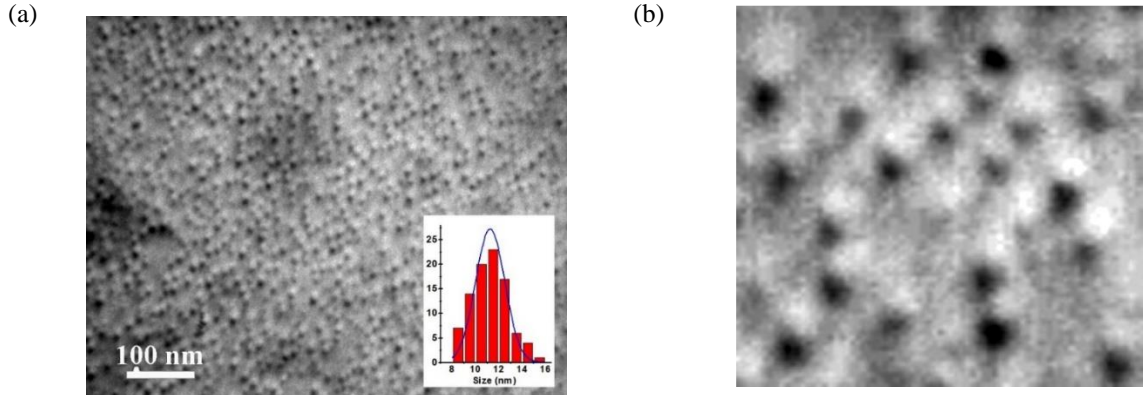

**Fig. 10.** A bright field TEM image of a similar InAs QD layer without the top AlAs, GaAs layers present in the actual sample used in our study. (a) A larger scan is shown on the left with estimated plot of size distributions. (b) A magnified version having 128 nm x128 nm scan size is given on the right.

The single crystal sample used in our measurements is exactly similar to one of those used by Vdovin *et al*. [8-10]. The layer structure, which was grown in University of Nottingham, consists of one layer of InAs self-assembled quantum dots (QDs) embedded within a GaAs/AlAs/InAs QDs/AlAs/GaAs p-i-n diode. The epilayers were grown by Molecular Beam Epitaxy (MBE) using solid sources on a highly doped (100) n$^+$ GaAs substrate. The p-type and n-type layers were achieved using Be impurities and Si, respectively. After desorption of the oxide layer from the substrate surface, a 1.0 µm heavily n-doped (Si = $4\times10^{18}$ /cm$^3$) GaAs buffer layer was grown at a temperature of 600 °C with at a growth rate of 1 micron/hour. This was followed by a 100 nm n-doped (Si = $2\times10^{16}$/cm$^3$) GaAs layer, a 100 nm undoped GaAs spacer layer, and two intrinsic AlAs quantum barriers of thicknesses of 5.1 nm each having a 1.8 monolayer of InAs QDs with areal density ~$1\times10^{11}$/cm$^2$ inserted between them. Then a 60 nm undoped GaAs layer completed the intrinsic region. Finally, a 0.51 µm heavily p-doped (Be = $2\times10^{18}$/cm$^3$) GaAs layer was deposited. The QDs were formed at a growth temperature of 520 °C using the well-known

Stranski–Krastanov growth mode. In order to determine the transition from a two dimensional to a three-dimensional growth mode, the reflection high-energy electron-diffraction (RHEED) pattern was monitored in a separate growth run. Bright field Transmission Electron Microscopy (TEM) image of the QDs without the capping layers are shown in Fig. 10. Surface density of these InAs QDs were ~1.0 x $10^{11}$/cm$^2$. Average spatial (x-y) extent of these QDs were around 11± 2 nm. The average z dimension of 1.6 nm was measured using Atomic Force Microscope (AFM).

To carry out the two-terminal electrical measurements described in this paper, samples were processed into circular mesas of diameter 200 μm. Ohmic contacts were prepared with Au/Ge/Ni alloying on the n$^+$ GaAs substrate and with Au-Zn alloying for the top p$^+$ GaAs layer of the heterostructure. Top contact metallization was ring-shaped (ring's diameter is around 25 μm) allowing normal incidence photoexcitation of the sample as shown in Fig. 1a.

For low temperature measurements, we placed our sample on a customized copper holder inside a closed-cycle cryostat CS-204S-DMX-20 from Advance Research Systems. Temperature of the cryostat was controlled with a Lakeshore (Model-340) temperature controller. The sample was illuminated from the top p-GaAs side using an Acton Research SP2555 monochromator having a 0.5-m focal length along with a 1000-W quartz-tungsten-halogen lamp from Newport as non-coherent light source or with a 633 nm and/or 488 nm laser as described. For photocapacitance measurements ($C_{photo}$ = dQ/dV), we used Agilent's E4980A LCR meter with small signal rms voltage of 30 mV with a frequency of 10 kHz. A simple series equivalent circuit of capacitance (C) and conductance (G) in parallel was used to extract these parameters. Spectral response of the lamp-monochromator combination is reasonably smooth and changes slowly and monotonically within the wavelength ranges used in our experiments. So Lambda-square corrections [60] were not used while plotting these spectra.

**Acknowledgements**

SD acknowledges the Science and Engineering Board (SERB) of Department of Science and Technology (DST), India (Grants # DIA/2018/000029, CRG/2019/000412 and SR/NM/TP13/2016) for supports. AB and MKS are thankful to DST, India for Inspire Ph.D Fellowship and IISER-Pune for Integrated Ph.D Fellowship, respectively. AB also received support from the Newton-Bhabha Ph.D Placement Programme of DST, India and British Council, UK. We thankfully acknowledge Rafeeque Mavoor of IISER-Pune's Media Centre for the illustration included in Fig. 9.

**Data and materials availability**

All data and materials used in the analyses will be available to support the findings of this study and/or for purposes of reproducing and/or extending these analyses from the corresponding author upon reasonable requests.

**CRediT authorship contribution statement**

SD along with AB, MKS had planned and designed the experiments on the sample grown and fabricated by MAH, MH. AB and MKS had collected the data with equal contributions. SD had analyzed and interpreted the data and wrote the paper in consultation with others.

**Declaration of Competing Interest**

The authors declare the following financial interests/personal relationships which may be considered as potential competing interests: Shouvik Datta reports financial support, administrative support, and travel were provided by Indian Institute of Science Education and Research Pune. Shouvik Datta reports a relationship with Indian Institute of Science Education and Research Pune that includes: employment, funding grants, and travel reimbursement. Authors had applied for a patent with International Application # PCT/IB2022/054992 and WO/2022/259078 pending to Indian Institute of Science Education and Research. Prof. Mohamed Henini is an Associate editor of this journal and a co-author of this paper as well.

# References

[1] P. Y. Yu, and M. Cardona, Section. 6.3, Fundamentals of Semiconductors, Springer, (1995).
[2] C. C. Escott, F. A. Zwanenburg, and A. Morello, Resonant tunneling features in quantum dots, Nanotechnology 21 (2010) 274018, doi:10.1088/0957-4484/21/27/274018.
[3] H. Mizuta, and T. Tanoue, The physics and applications of resonant tunneling diodes, Cambridge University Press, (1995).
[4] L. L. Chang, E. E. Mendez, and C. Tejedor, Resonant Tunneling in Semiconductors: Physics and Applications Springer, (1990).
[5] A. E. Belyaev, L. Eaves, P. C. Main, Á. Polimeni, S. T. Stoddart, and M. Henini, Capacitance Spectroscopy of Single-Barrier GaAs/AlAs/GaAS Structures Containing InAs Quantum Dots, Acta. Phys. Polonica A 94 (1998) 245-249, doi:10.12693/APhysPolA.94.245.
[6] S. Pal, C. Junggebauer, S. R. Valentin, P. Eickelmann, S. Scholz, A. Ludwig, and A. D. Wieck, Probing indirect exciton complexes in a quantum dot molecule via capacitance-voltage spectroscopy, Phys. Rev. B 94 (2016) 245311, doi:10.1103/PhysRevB.94.245311.
[7] P. A. Labud, A. Ludwig, A. D. Wieck, G. Bester, and D. Reuter, Direct Quantitative Electrical Measurement of Many-Body Interactions in Exciton Complexes in InAs Quantum Dots, Phys. Rev. Lett. 112 (2014) 046803, doi:10.1103/PhysRevLett.112.046803.
[8] E. E. Vdovin, M. Ashdown, A. Patanè, L. Eaves, R. P. Campion, Yu. N. Khanin, M. Henini, and O. Makarovsky, Quantum oscillations in the photocurrent of GaAs/AlAs p-i-n diodes, Phys. Rev. B 89 (2014) 205305, doi:10.1103/PhysRevB.89.205305.
[9] E. E. Vdovin and Y. N. Khanin,, Effect of the radiation power on the modification of oscillations of the photocurrent in single-barrier p−i−n GaAs/AlAs/GaAs heterostructures with InAs quantum dots, JETP Lett. 113 (2021) 586-591, doi:10.31857/S123456782109007X.
[10] Y. N. Khanin, E. E. Vdovin and S. V. Morozova, Strong effect of the wavelength of light on quantum oscillations of the photocurrent and their resonant tunneling nature in p−i−n GaAs/AlAs heterostructures, JETP Lett. 114 (2021) 332-336, doi:10.31857/S1234567821180063.
[11] B. Skinner and B. I. Shklovskii, Anomalously large capacitance of a plane capacitor with a two-dimensional electron gas, Phys. Rev. B 82 (2010) 155111, doi:10.1103/PhysRevB.82.155111.
[12] J. P. Eisenstein, L. N. Pfeiffer, and K. W. West, Negative compressibility of interacting two-dimensional electron and quasiparticle gases, Phys. Rev. Lett. 68 (1992) 674, doi:10.1103/PhysRevLett.68.674.
[13] J. P. Eisenstein, L. N. Pfeiffer, and K. W. West, Compressibility of the two-dimensional electron gas: Measurements of the zero-field exchange energy and fractional quantum Hall gap, Phys. Rev. B 50 (1994) 1760, doi:10.1103/PhysRevB.50.1760.
[14] A. Bhunia, M. K. Singh, Y. G. Gobato, M. Henini, and S. Datta, Experimental Detection and Control of Trions and Fermi-Edge Singularity in Single-Barrier GaAs/AlAs/GaAs Heterostructures Using Photocapacitance Spectroscopy, Phys. Rev. Appl. 10 (2018) 044043, doi: 10.1103/PhysRevApplied.10.044043.

[15] A. Bhunia, M. K. Singh, Y. G. Gobato, M. Henini, and S. Datta, Experimental evidences of quantum confined 2D indirect excitons in single barrier GaAs/AlAs/GaAs heterostructure using photocapacitance at room temperature J. Appl. Phys. 123 (2018) 044305, doi: 10.1063/1.5007820.

[16] A. Bhunia, K. Bansal, M. Henini, M. S. Alshammari, and S. Datta, Negative activation energy and dielectric signatures of excitons and excitonic Mott transitions in quantum confined laser structures, J. Appl. Phys. 120 (2016) 144304, doi:10.1063/1.4964850.

[17] M. Buttiker, Coherent and sequential tunneling in series barriers, IBM J. Res. Dev. 32 (1988) 63-75, doi:10.1147/rd.321.0063.

[18] Y. N. Joglekar, A. V. Balatsky, and S. D. Sarma, Wigner supersolid of excitons in electron-hole bilayers, Phys. Rev. B 74 (2006) 233302, doi:10.1103/PhysRevB.74.233302.

[19] L. V. Butov, C. W. Lai, A. L. Ivanov, A. Gossard, and D. S. Chemla, Towards Bose–Einstein condensation of excitons in potential traps, Nature 417 (2002) 47-52, doi:10.1038/417047a.

[20] J. P. Eisenstein, A. H. MacDonald, Bose–Einstein condensation of excitons in bilayer electron systems, Nature 432 (2004) 691-694, doi:10.1038/nature03081.

[21] M. M. Fogler, L. V. Butov. K. S. Novoselov, High-temperature superfluidity with indirect excitons in van der Waals heterostructures, Nat. Commun. 5 (2014) 4555, doi:10.1038/ncomms5555.

[22] A. Kogar, M. S. Rak, S. Vig, A. A. Husain, F. Flicker, Y. Il Joe, L. Venema, G. J. MacDougall, T. C. Chiang, E. Fradkin, J. van Wezel, and P. Abbamonte, Signatures of exciton condensation in a transition metal dichalcogenide, Science 358 (2017) 1314-1317, doi:10.1126/science.aam643.

[23] J. P. Eisenstein, L. N. Pfeiffer, and K. W. West, Precursors to Exciton Condensation in Quantum Hall Bilayers, Phys. Rev. Lett. 123 (2019) 066802, doi:10.1103/PhysRevLett.123.066802.

[24] Z. Wang, D. A. Rhodes, K. Watanabe, T. Taniguchi, J. C. Hone, J. Shan, and K. F. Mak, Evidence of high-temperature exciton condensation in two-dimensional atomic double layers, Nature 574 (2019) 76-80, doi:10.1038/s41586-019-1591-7.

[25] D. W. Snoke, Indirect Excitons in Coupled Quantum Wells, Final Report, DOE grant DE-FG02-99ER45780, (2014), doi:10.2172/1141286.

[26] D. W. Snoke, When should we say we have observed Bose condensation of excitons?, Phys. Status Solidi B 238 (2003) 389-396, doi:10.1002/pssb.200303151.

[27] M. Combescot, O. Betbeder-Matibet, and R. Combescot, Bose-Einstein Condensation in Semiconductors: The Key Role of Dark Excitons, Phys. Rev. Lett. 99 (2007) 176403, doi:10.1103/PhysRevLett.99.176403.

[28] M. Combescot, R. Combescot, and F. Dubin, Bose–Einstein condensation and indirect excitons: a review, Rep. Prog. Phys. 80 (2017) 066501, doi:10.1088/1361-6633/aa50e3.

[29] F. Capasso, K. Mohamed, and A. Y. Cho, Resonant Tunneling Through Double Barriers, Perpendicular Quantum Transport Phenomena in Superlattices, and Their Device Applications, IEEE J. of Quant. Electron. QE-22 (1986) 1853-1869, doi:10.1109/JQE.1986.1073171.

[30] T. Lahaye, C. Menotti, L. Santos, M. Lewenstein, and T. Pfau, The physics of dipolar bosonic quantum gases, Rep. Prog. Phys. 72 (2009) 126401, doi:10.1088/0034-4885/72/12/126401.

[31] Z. Hadzibabic, and J. Dalibard, Two-dimensional Bose fluids: An atomic physics perspective, Rivista del Nuovo Cimento 34 (2011) 389-434, doi:10.1393/ncr/i2011-10066-3.

[32] O. L. Berman, and R. Y. Kezerashvili, High-temperature superfluidity of the two-component Bose gas in a transition metal dichalcogenide bilayer, Phys. Rev. B 93 (2016) 245410, doi:10.1103/PhysRevB.93.245410.

[33] D. Cavalcanti, Split, but still attached, Science 360 (2018) 376-377, doi:10.1126/science.aat459.

[34] L. Heaney, J. Anders, D. Kaszlikowski, and Vlatko Vedral, Spatial entanglement from off-diagonal long-range order in a Bose-Einstein condensate, Phys. Rev. A 76 (2007) 053605, doi:10.1103/PhysRevA.76.053605.

[35] A. Zrenner, E. Beham, S. Stufler, F. Findeis, M. Bichler, and G. Abstreiter, Coherent properties of a two-level system based on a quantum-dot photodiode, Nature 418 (2002) 612-614, doi:10.1038/nature00912.

[36] X. Li, Y. Wu, D. Steel, D. Gammon, T. H. Stievater, D. S. Katzer, D. Park, C. Piermarocchi, and L. J. Sham, An All-Optical Quantum Gate in a Semiconductor Quantum Dot, Science 301 (2003) 809-811, doi:10.1126/science.1083800.

[37] N. H. Bonadeo, J. Erland, D. Gammon, D. Park, D. S. Katzer, D. G. Steel, Coherent Optical Control of the Quantum State of a Single Quantum Dot, Science 282 (1998) 1473-1476, doi:10.1126/science.282.5393.14.

[38] T. H. Stievater, X. Li, D. G. Steel, D. Gammon, D. S. Katzer, D. Park, C. Piermarocchi, and L. J. Sham, Rabi Oscillations of Excitons in Single Quantum Dots, Phys. Rev. Lett. 87 (2001) 133603, doi:10.1103/PhysRevLett.87.133603.

[39] H. Kamada, H. Gotoh, J. Temmyo, T. Takagahara, and H. Ando, Exciton Rabi Oscillation in a Single Quantum Dot, Phys. Rev. Lett. 87 (2001) 246401, doi:10.1103/PhysRevLett.87.246401.

[40] M. R. Andrews, Observation of Interference Between Two Bose Condensates, Science 275 (1997) 637-641, doi:10.1126/science.275.5300.6.

[41] Y. Lin, J. P. Gaebler, F. Reiter, T. R. Tan, R. Bowler, A. S. Sørensen, D. Leibfried, and D. J. Wineland, Dissipative production of a maximally entangled steady state of two quantum bits, Nature 504 (2013) 415-418, doi:10.1038/nature12801.

[42] Z. Wang, W. Wu, and Jin Wang, Steady-state entanglement and coherence of two coupled qubits in equilibrium and nonequilibrium environments, Phys. Rev. A 99 (2019) 042320, doi:10.1103/PhysRevA.99.042320.

[43] J. I. Cirac, and P. Zoller, Quantum Computations with Cold Trapped Ions, Phys. Rev. Lett. 74 (1995) 4091, doi:10.1103/PhysRevLett.74.4091.

[44] J. I. Cirac, M. Lewenstein, K. Mølmer, K. and P. Zoller, Quantum superposition states of Bose-Einstein condensates, Phys. Rev. A 57 (1998) 1208, doi:10.1103/PhysRevA.57.1208.

[45] S. M. H. Halataei, and A. J. Leggett, Tunnel Splitting in Asymmetric Double Well Potentials: An Improved WKB Calculation, arxiv:1703.05758 (2017), doi:10.48550/arXiv.1703.05758.

[46] L. Q. Lambert, A. Compaan, and I. D. Abella, Effects of Nearly Degenerate States on Photon-Echo Behavior, Phys. Rev. A 4 (1971) 2022, doi:10.1103/PhysRevA.4.2022.

[47] A. J. Leggett, Macroscopic Quantum Systems and the Quantum Theory of Measurement, Prog. Theor. Phys. Suppl. 69 (1980) 80-100, doi:10.1143/PTP.69.80.

[48] J. Preskill, Lecture Notes for Physics 229, Quantum Information and Computation, California Institute of Technology, (1998).

[49] T. Byrnes, Kai Wen, and Yoshihisa Yamamoto, Macroscopic quantum computation using Bose-Einstein condensates, Phys. Rev. A 85 (2012) 040306R, doi:10.1103/PhysRevA.85.040306.

[50] S. Ghosh & T. C. H. Liew, Quantum computing with exciton-polariton condensates, npj Quantum Information, 6 (2020) 16, doi:10.1038/s41534-020-0244-x.

[51] S. Rosenblum, P. Reinhold, M. Mirrahimi, L. Jiang, L. Frunzio, and R. J. Schoelkopf, Fault-tolerant detection of a quantum error, Science 361 (2018) 266-271, doi:10.1126/science.aat3996.

[52] B. M. Terhal, J. Conrad, and C. Vuillot, Towards scalable bosonic quantum error correction, Quantum Sci. Technol. 5 (2020) 043001, doi:10.1088/2058-9565/ab98a5.

[53] J. M. Gertler, B. Baker, J. Li, S. Shirol, J. Koch and C. Wang, Protecting a bosonic qubit with autonomous quantum error correction, Nature, 590 (2021) 243-248, doi:10.1038/s41586-021-03257-0.

[54] D. P. DiVincenzo, The Physical Implementation of Quantum Computation, Fortschritte der Physik. 48 (2000) 771-783. doi:10.1002/1521-3978(200009)48:9/113.0.CO;2-E.

[55] J. Wang, P. Nie, X. Li, H. Zuo, B. Fauque, Z. Zhua and K. Behnia, Critical point for Bose–Einstein condensation of excitons in graphite, Proc. Natl. Acad. Sci, 117 (2020) 30215–30219, doi:10.1073/pnas.2012811117.

[56] D. Wang, N. Luo, W. Duan, and X. Zou, High-Temperature Excitonic Bose-Einstein Condensate in Centrosymmetric Two-Dimensional Semiconductors, J. Phys. Chem. Lett.12 (2021) 5479−5485, doi:10.1021/acs.jpclett.1c01370

[57] H. Guo, X. Zhang and G. Lu, Tuning moiré excitons in Janus heterobilayers for high-temperature Bose-Einstein condensation, Sci. Adv. 8 (2022) eabp9757, doi:10.1126/sciadv.abp9757.

[58] Y. Zhang, B. Hou, Y. Wu, Y. Chen, Y. Xia, H. Mei, M. Kong, L. Peng, H. Shao, J. Cao, W. Liu, H. Zhu, H. Zhang, Towards high-temperature electron-hole condensate phases in monolayer tetrels metal halides: Ultra-long excitonic lifetimes, phase diagram and exciton dynamics, Mat. Today. Phys. 22 (2022) 100604. doi:10.1016/j.mtphys.2022.100604.

[59] Y. Morita, K. Yoshioka and M. Kuwata-Gonokami, Observation of Bose-Einstein condensates of excitons in a bulk semiconductor, Nat. Comm. 13 (2022) 5388, doi:10.1038/s41467-022-33103-4.

[60] I. Pelant, and J. Valenta, Section 2.7, Luminescence Spectroscopy of Semiconductors, Oxford University Press, New York, (2012 ).

# Supplementary Material

# 0D-2D Heterostructure for making very Large Quantum Registers using 'itinerant' Bose-Einstein Condensate of Excitons

Amit Bhunia[1], Mohit Kumar Singh[1], Maryam Al Huwayz[2,3], Mohamed Henini[2] and Shouvik Datta[1] *

[1]*Department of Physics, Indian Institute of Science Education and Research, Pune 411008, Maharashtra, India*
[2]*School of Physics and Astronomy, University of Nottingham, Nottingham NG7 2RD, UK,*
[3]*Physics Department, Faculty of science, Princess Nourah Bint Abdulrahman University, Riyadh, Saudi Arabia.*

***Email:*** [shouvik@iiserpune.ac.in](shouvik@iiserpune.ac.in), [Ppzmh@exmail.nottingham.ac.uk](Ppzmh@exmail.nottingham.ac.uk)

**Additional justifications for excitonic BEC**

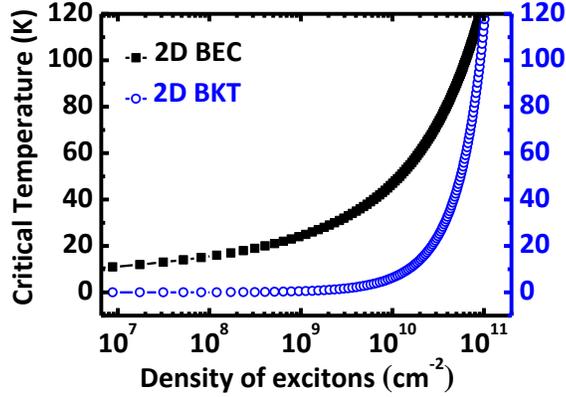

**Fig. S1.** Simulated plots of critical temperature of BEC and BKT transitions as a function of 2D density of excitons. This supports excitonic BEC around 100 K for indirect excitons with areal density ~$10^{11}$/cm² in this quasi 2D heterostructure of III-V semiconductors.

We will now discuss additional details on why this excitonic BEC is observable at these parameter ranges along with the measured exciton densities. An estimate of the critical surface density of 2D bosons as the order parameter [S1] for BEC is given by $N_{BEC}^{2D} = 2\left(\frac{1}{\lambda_{Th}}\right)^2 ln\left(\frac{L}{\lambda_{Th}}\right)$, where $\lambda_{Th}$ is the thermal de-Broglie wavelength for indirect excitons in this case, L is the lateral extent (~200 μm) of the 2DEG being probed in x-y plane. As mentioned in the manuscript, here we have used $m_{exciton} \sim m^*$ for localized excitons where holes are confined within InAs QDs under reverse bias. This is because thermal degrees of freedom of these 0D-2D indirect excitons are mostly associated with lighter electrons in GaAs 2DEG. This $m^*$ is 0.058$m_0$ where $m_0$ is the free electron mass, Using this value, we obtained the threshold density for 2D excitons as $N_{BEC}^{2D} \sim 1.52 \times 10^{11}$/cm². Even this estimate is close to the measured dipolar density $\sigma_{ph}$ of 0D-2D indirect excitons at 10 K. So it is likely that actual BEC coherence length can be larger than spacing between the neighboring InAs QDs. As a result, wave functions of these quantum coupled, 0D-2D indirect excitons in fact begin to overlap below 100 K. Interestingly, these quantum coherent oscillations slowly decohere and gradually vanish with enhanced charge accumulations under higher DC bias voltages and also at higher photo excitation intensities. They certainly vanish with increasing temperatures above ~100 K [Fig. 4].

Critical density of BEC in this quasi-2D system can also be estimated using [S2] $n_{EX} \sim N_{BEC}^{2D} = -\left(\frac{1}{\lambda_{Th}^2(T_C^{BEC})}\right) ln\left(1 - \exp(-\frac{|\epsilon_0|}{k_B T_C^{BEC}})\right)$. In Fig. S1, we plot the critical temperature $(T_C^{BEC})$ for attaining excitonic BEC as a function $n_{EX}$, where $|\epsilon_0| = 6.8$ meV is assumed to be the excitonic binding energy [S2, S3] in such III-V material systems. We also plot the following Berezinskii-Kosterlitz-Thouless (BKT) [S4] transition temperature $(T_C^{BKT})$ in Fig. S1, where $a_0$ is the Bohr exciton radius. $a_0$ is assumed to be $\approx 10$ nm for $|\epsilon_0| = 6.8$ meV and $k_B T_C^{BKT} = \frac{(\hbar^2/2m^*)4\pi n_{EX}}{ln\left(ln\left(\frac{1}{n_{EX} a_0^2}\right)\right)}$. We see that $T_C^{BKT} < T_C^{BEC}$, but both are still around 100 K or more for $n_{EX} \sim \sigma_{Ph} \sim 10^{11}$/cm². This temperature of ~100 K is already well within the operational range of liquid nitrogen based cryogenics being used already for some commercial supercomputers.

Moreover, the areal density of these 0D-2D indirect excitons as BEC order parameter is also the limiting value of first order spatial correlation function $g_1(r)$ such that [S5,S6] $n_{EX} \sim \sigma_{Ph} \equiv \lim_{r \to \infty}[g_1(r)]$. We already estimated this as $n_{EX} \sim \sigma_{Ph} \sim 10^{11}$/cm² even when $r \to 200$ $\mu m$. So, we expect the interacting Bose gas of dipolar, indirect excitons to exhibit quasi long-range order over such a macroscopically large area. Although we have not yet directly measured this $g_1(r)$ using localized photocapacitance measurements, which is currently beyond the scope of this study. However, we demonstrated that photocapacitance of two spatially separated ensembles of excitons generated even with white light can coherently interfere when light spots overlap (Section 7, Fig. 5). Any precise identification of BEC or BKT phase of these excitons and knowing the exact fraction of dark to bright excitons is also currently beyond the scope of this study.

**Additional supporting results on the differences between photocapacitance and photoconductance measurements.**

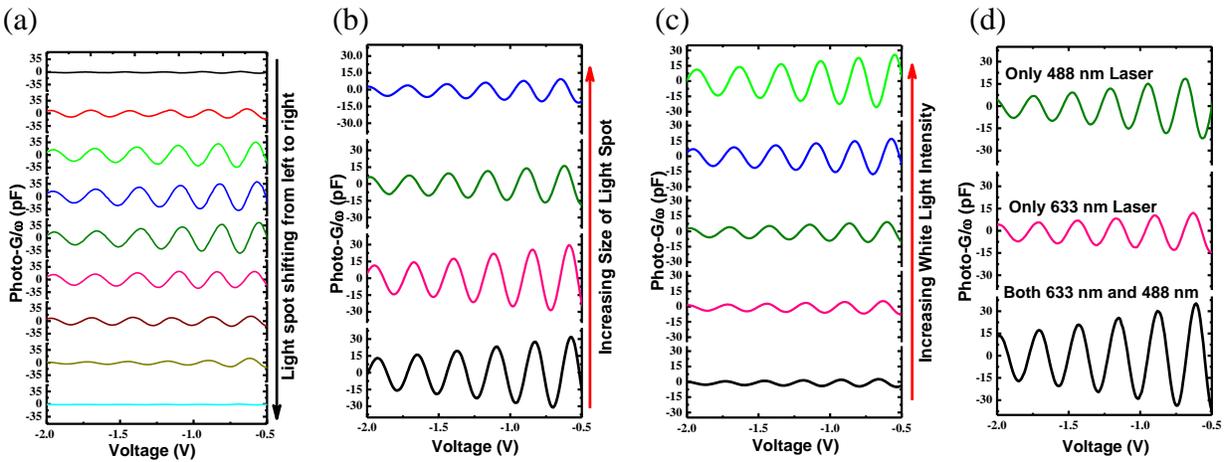

**Fig. S2.** Unlike photocapacitance [Fig. 5], we do not see any quantum interference beats in photo-G/ω. (a) Photo-G/ω does not show beat like features when the laser spot is moved as described in Fig. 5(a). (b) No beat formation in photo-G/ω as the laser spot was increased gradually beyond the periphery of the ring. (c) No beat formation even with white light. (d) Quantum interference beats were absent when two separate, uncorrelated, CW laser beams of 633 nm and 488 nm are used to photo excite the sample simultaneously. Therefore, unlike photocapacitance, photo-G/ω is not specific to the phase coherent, collective quantum properties of dipolar excitons at the 0D-2D heterojunction.

**Additional supporting results on splitting of photocapacitance spectra**

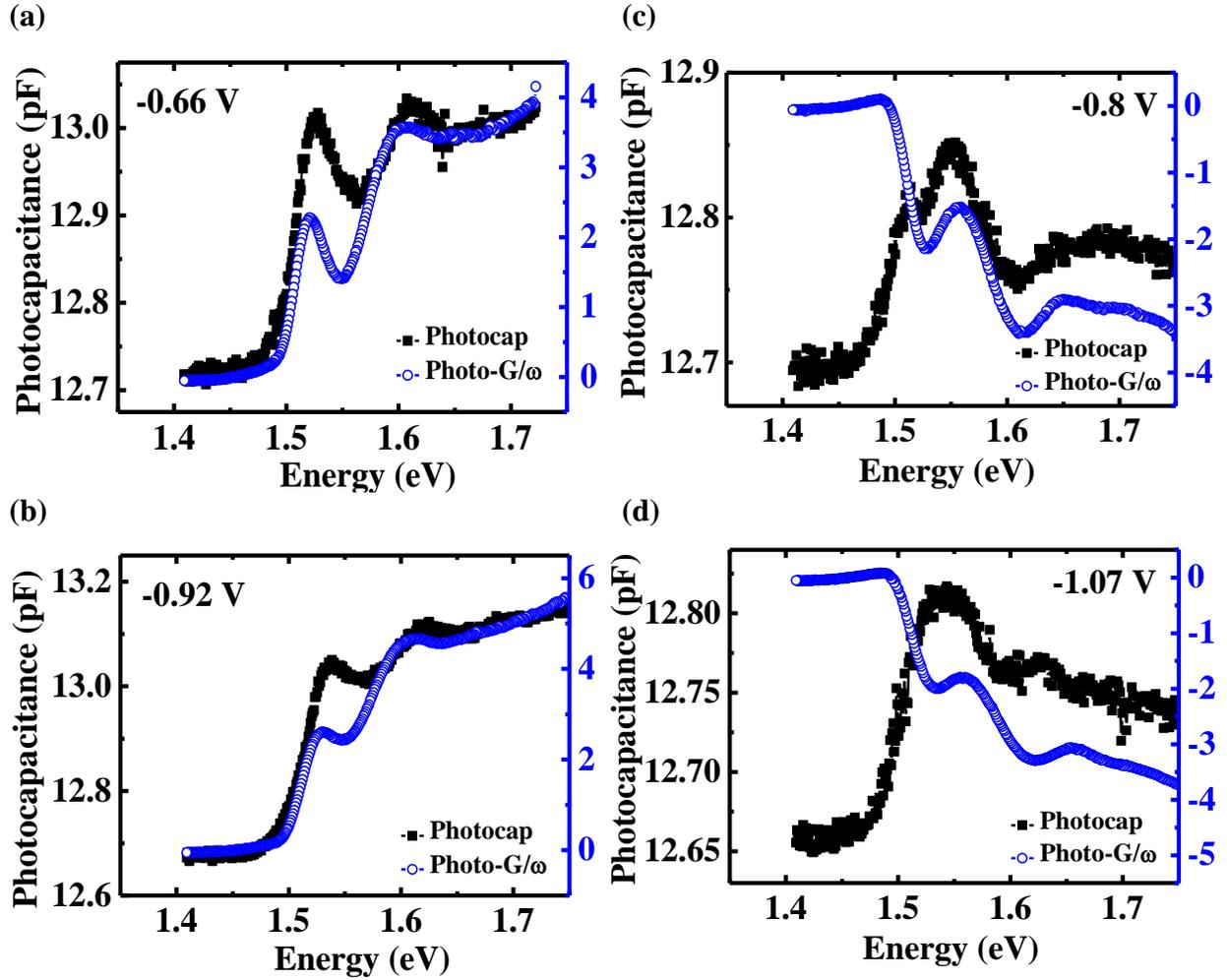

**Fig. S3.** Presence and absence in sharply split excitonic transitions in photocapacitance spectra at other bias voltages at 10 K. (a) and (b) Photocapacitance and photo-G/ω spectra under biases of successive maximum of photocapacitance oscillations. Corresponding biases are -0.66 V, -0.92 V respectively from Fig. 1c. (c) and (d) Photocapacitance and photo-G/ω spectra under biases of successive minimum of photocapacitance oscillations. Corresponding biases are -0.80 V, -1.07 V respectively from Fig. 1c. Therefore, unlike photocapacitance, photo-G/ω cannot sense the tunneling induced quantum coupling and reorganization of the 0D-2D indirect excitons.

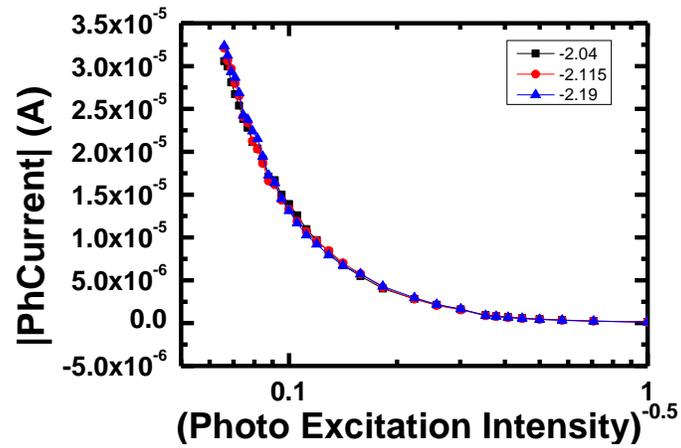

**Fig. S4.** Measured dc-photocurrent increases monotonically with increasing photoexcitation intensity.

**Additional Supplementary References**

[S1] W. Ketterle, and N. J. V. Druten, Bose-Einstein condensation of a finite number of particles trapped in one or three dimensions, Phys. Rev. A 54 (1996) 656, doi:10.1103/PhysRevA.54.656.
[S2] J. F.Jan, and Y. C. Lee, Bose-Einstein condensation of excitons in two dimensions, Phys. Rev. B 58 (1998) 1714R, doi:10.1103/PhysRevB.58.R1714.
[S3] G. D. Gilliland, A. Antonelli, D. J. Wolford, K. K. Bajaj, J. Klem, and J. A. Bradley, Direct measurement of heavy-hole exciton transport in type-II GaAs/AlAs superlattices, Phys. Rev. Lett. 71 (1993) 3717, doi: 10.1103/PhysRevLett.71.3717.
[S4] D. S. Fisher, and P. C Hohenberg, Dilute Bose gas in two dimensions, Phys. Rev. B 37 (1988) 4936, doi: 10.1103/PhysRevB.37.4936.
[S5] O. Penrose, and L. Onsager, Bose-Einstein Condensation and Liquid Helium, Phys. Rev. 104 (1956) 576, doi: 10.1103/PhysRev.104.57.
[S6] V. I. Yukalov, Basics of Bose-Einstein condensation, Phys. Part. Nucl. 42 (2011) 460-513, doi:10.1134/S1063779611030063.